\documentclass[]{JHEP3}

\def\cA{\mathcal{A}}%%%%%%%%%% calligraphy

\def\cC{\mathcal{C}}
\def\cD{\mathcal{D}}
\def\cE{\mathcal{E}}
\def\cF{\mathcal{F}}
\def\cG{\mathcal{G}}
\def\ceH{\mathcal{H}}
\def\cI{\mathcal{I}}

\def\cM{\mathcal{M}}
\def\cN{\mathcal{N}}
\def\cO{\mathcal{O}}

\def\cR{\mathcal{R}}
\def\cS{\mathcal{S}}
\def\cT{\mathcal{T}}
\def\cU{\mathcal{U}}

\def\xcA{\mathscr{A}}%%%%%%%%%% extra calligraphy

\def\xcD{\mathscr{D}}

\def\xcG{\mathscr{G}}

\def\xcO{\mathscr{O}}

\def\xcT{\mathscr{T}}
\def\xcU{\mathscr{U}}

\def\bC{{\mathbb{C}}}

\def\bR{{\mathbb{R}}}

\def\bZ{{\mathbb{Z}}}

\def\a{\alpha}%%%%%%%%%% Greek
\def\b{\beta}
\def\g{\gamma}

\def\d{\delta}
\def\D{\Delta} 
\def\ep{\epsilon}

\def\th{\theta}

\def\la{\lambda}
\def\La{\Lambda}
\def\om{\omega}
\def\Om{\Omega}

\def\si{\sigma}
\def\Si{\Sigma}

\def\ggt{\gt{g}}

\newcommand{\sfd}{{\mathsf d}}

\newcommand{\sfi}{{\mathsf i}}

\newcommand{\sfk}{{\mathsf k}}

\def\bd{{\dot \b}}

\def\x{\times}%%%%%%%%%% multiplication and actions
\def\ox{\otimes}

\def\lact{\vartriangleright}

\def\p{\partial}%%%%%%%%%%%%%% derivatives

\def\sign{{\rm sign}}

\def\dim{{\rm dim}}%%%%%%%%% linear spaces

\def\Hol{\tx{Hol}}

\def\Cliff{\tx{Cliff}}%%%%%%%%%%% algebras

\def\det{{\rm det}}%%%%%%%%% matrices

\def\Mat{\textrm{Mat}}

\def\tr{{\rm tr}}
\def\Tr{{\rm Tr}}

\def\qgt{\cU_q(\ggt)}

\newcommand{\faff}[1]{P^{\sfk}_{+}(#1)}

\newcommand{\lb}{\left(}%%%%%%%%%%%%%%% brackets
\newcommand{\rb}{\right)}

\newcommand{\lan}{\langle}
\newcommand{\ran}{\rangle}

\newtheorem{defn}{Definition}[section]

\newtheorem{exer}{Exercise}[section]
\newtheorem{soln}{Solution}[section]

\newtheorem{rmq}{Remark}[section]

\newcommand{\beq}{\begin{equation}}%%%%%%%%%%%% environments
\newcommand{\eeq}{\end{equation}}
\newcommand{\beqa}{\begin{eqnarray}}
\newcommand{\eeqa}{\end{eqnarray}}
\newcommand{\begt}{\begin{gather}}
\newcommand{\bal}{\begin{align}}
\newcommand{\eal}{\end{align}}

\newcommand{\barr}{\begin{array}}
\newcommand{\earr}{\end{array}}
\newcommand{\ben}{\begin{enumerate}}
\newcommand{\een}{\end{enumerate}}
\newcommand{\bit}{\begin{itemize}}
\newcommand{\eit}{\end{itemize}}
\newcommand{\bdef}{\begin{defn}}
\newcommand{\eedef}{\end{defn}}
\newcommand{\bermq}{\begin{rmq}}
\newcommand{\eermq}{\end{rmq}}
\newcommand{\bex}{\begin{exer}}
\newcommand{\eex}{\end{exer}}
\newcommand{\besol}{\begin{soln}}
\newcommand{\eesol}{\end{soln}}
\newcommand{\tx}[1]{\textrm{#1}} 
 
\newcommand{\gt}[1]{\mathfrak{#1}}

\def\must{\stackrel{!}{=}}%%%%%%%%%%% varia

\def\1{\mathbb{I}}
\def\2{[2]_q}

\newcommand{\non}{\nonumber\\}
\newcommand{\nn}{\nonumber}

\newcommand{\nl}{\newline}
\newcommand{\qq}{\begin{eqnarray}}

\newcommand{\ee}{{\rm e}}
\newcommand{\qqq}{\end{eqnarray}}

\newcommand{\rep}{representation }

\newcommand{\irrep}{irreducible representation }

\newcommand{\wrt}{with respect to }

\def\gvee{g^{\vee}}

\def\qU2{\cU_q (\mathfrak{su}(2))}

\def\too{\longmapsto}

\def\kmg{\widehat{\mathfrak{g}}_{\sfk}}

\def\resu2{\textrm{REA}_q(\mathfrak{su}(2))}

\def\Cv{\v{C}}
\def\vC{\check{C}}

\newcommand{\unl}[1]{\underline{#1}}

\def\bd1{{\boldsymbol{1}}}
\def\brd0{{\boldsymbol{0}}}

\usepackage{latexsym,amsfonts,amsmath,amssymb,mathrsfs,pifont}
\usepackage{indentfirst}

\title{Non-commutative Weitzenb\"ock geometry,\\ gerbe modules, and WZW branes}
\author{Andreas Recknagel$^{\,a}$\ \  and\ \  Rafa\l\ R. Suszek$^{\,b}$ \\
${}^a\,$King's College London, Department of Mathematics, Strand,
London WC2R 2LS, UK\\
${}^b\,$Laboratoire de Physique, \'Ecole Normale Sup\'erieure de Lyon, 
46 All\'ee d'Italie, \\ \phantom{${}^b\,$}F-69364 Lyon, France\\ %\vskip-9pt
$\ $\email{andreas.recknagel@kcl.ac.uk, rafal.suszek@ens-lyon.fr}
}

\abstract{We study the non-commutative matrix model which arises as the low-energy
effective action of open strings in WZW models. We re-derive this fuzzy effective
gauge dynamics in two different ways, without recourse to conformal field
theory: The first method starts from a linearised version of the WZW $\si$-model,
which is classically equivalent to an action of the Schild type, which in
turn can be quantised in a natural way to yield the matrix model. The second
method relies on purely geometric symmetry principles -- albeit within the
non-commutative spectral geometry that is provided by the boundary CFT data:
we show that imposing invariance under extended gauge transformations singles
out the string-theoretic action up to the relevant order in the gauge field.
The extension of ordinary gauge transformations by tangential shifts is motivated
by the gerbe structure underlying the classical WZW model and standard within
Weitzenb\"ock geometry -- which is a natural reformulation of geometry to
use when describing strings in targets with torsion. 
}

\keywords{Conformal field theory, WZW models, D-branes on group manifolds;
matrix models; Schild action; curved non-commutative geometry, spectral triples; 
bundle gerbes, gerbe modules; torsion, teleparallelism}

\begin{document}

\section{Introduction}\label{sec:intro}

\noindent
Branes have become a central object in string theory, and one nice feature
is that they can be studied from various points of view, including target
space geometry as well as the world-sheet CFT. String backgrounds where both
descriptions are available, among them group targets and, to some extent,
Calabi--Yau manifolds, offer particularly interesting insights. Here we will
concentrate on WZW models, which are non-trivial in that their targets are
curved and carry torsion, but which are still highly symmetric and hence
tractable. The study of open-string dynamics has led to a more prominent
r\^ole of non-commutative geometry, at a more practical level through the analysis
of effective actions, and at a more fundamental level when discussing the
true nature of geometry in string theory.  Again, WZW models offer an ideal
playground to test ideas -- as we will try here by emphasising the r\^ole of
torsion within the NCG picture. 

The object we will be concerned with in this article is the low-energy effective
action of open strings in WZW models with compact simple and simply connected
target Lie group $\,G$,\, as derived in \cite{ars1,ars2}; this can be written
in a concise form as a matrix model for $\,X\in\Mat(N;\bC)$,
\qq\label{SARSX}
\cS_{\rm ARS}[X]=\tr\,\lb-\frac{1}{4}\,[\,X_a,X_b\,]\,[\,X_a,X_b\,]+
\frac{\sfi}{3}\,f_{abc}\,X_a\,[\,X_b,X_c\,]+\mu\cdot\bd1_N\rb\,,
\qqq
where $\,\mu\,$ is a constant and $\,a=1,2,\dots,d\equiv\dim G$.\, A less
compact, but ultimately more geometric and physical reformulation in terms
of dynamic variables $\,A_a\in\Mat(N;\bC)\,$ is obtained by introducing the
objects
\qq\label{A2cov}
X_a=:Y_a+A_a
\qqq
where the $\,N\times N\,$ matrices $\,Y_a\,$ furnish a representation of
the Lie algebra $\,\ggt\,$ of $\,G$,
\qq\nn
[\,Y_a,Y_b\,]=\sfi\,f_{abc}\,Y_c\,.
\qqq
From the CFT point of view, the $\,Y_a\,$ are induced by the action of the
horizontal Lie algebra $\,\ggt\,$ of the current symmetry algebra $\,\kmg\,$
of the boundary theory on CFT primaries (see the Appendix). From the point
of view of non-commutative geometry, the redefinition \eqref{A2cov} marks
$\,X_a\,$ as a so-called covariant coordinate, an object composed of an inner
derivation $\,Y_a\,$ (the rigid NCG background) and a gauge field $\,A_a\,$
(a fluctuation of the background), which on the whole is covariant under
the action of the unitary gauge group.

In terms of the gauge field $\,A_a\,$ on the brane, the action \eqref{SARSX}
indeed takes on a familiar form: It can be written as the sum
\qq\label{SARS}
\cS_{\rm ARS}[A]=\cS_{\rm YM}[A]+\cS_{\rm CS}[A]
\qqq
of a Yang--Mills and a Chern--Simons term
\qq\nn
\cS_{\rm YM}[A]=\frac{1}{4}\,\tr\,\bigl[\,F_{ab}(A)\,F_{ab}(A)\,\bigl]\,,
\qquad\cS_{\rm CS}[A]=-\frac{\sfi}{2}\,\tr\,\bigl[\,f_{abc}\,CS_{abc}(A)\,
\bigl]\,,
\qqq
with
\qq
F_{ab}(A)=\sfi\,[\,X_a,X_b\,]+f_{abc}\,X_c=\sfi\,[\,Y_a,A_b\,]-\sfi\,[\,
Y_b,A_a\,]+\sfi\,[\,A_a,A_b\,]+f_{abc}\,A_c\,, \label{FFzz} \\ \non
CS_{abc}(A)=A_a\,[\,Y_b,A_c\,]+\frac{2}{3}\,A_a\,A_b\,A_c-\frac{\sfi}{2}\,
f_{abd}\,A_d\,A_c\,.\nonumber
\qqq
One recognises that the $\,Y_a\,$ take over the r\^ole of derivatives on
flat D-brane world-volumes. In rewriting \eqref{SARSX} in this form, the
constant $\,\mu\,$ has been chosen proportional to the Casimir eigenvalue
$\,Y_a\,Y_a$.

In \cite{ars2}, this action was derived starting from the algebraic boundary
CFT description of untwisted maximally symmetric WZW branes, which are given by Cardy
boundary states labelled by elements $\,\la\in\faff{\ggt}\,$ of the
set of dominant integral affine weights of the Kac--Moody algebra $\,\kmg\,$
at level $\,\sfk$.\, The open-string states of interest are $\,A_a\,
j^a_{-1}\,\psi_l(x)$,\, where $\,\psi_l\,$ is a boundary field from the 
$\,\ggt$-multiplet associated with some $\,\kmg$-primary (so the $\,A_a\,$
are matrices of coefficients). As long as $\,0\lesssim\Vert\la\Vert\ll\sfk$,\,
these open-string states become massless in the low-energy (or decoupling)
limit
\qq\label{decouplinglimit}
\a'\longrightarrow 0\,,\qquad\qquad\sfk\longrightarrow\infty\,,\qquad\qquad
\sqrt{\a'\,\sfk}\equiv\ell\longrightarrow\infty\,.
\qqq
To deduce the effective action for those strings, one needs to exploit detailed
knowledge of WZW conformal field theories to compute all CFT $\,n$-point
correlation functions of these physical open-string states (dressed by ghosts
as usual) up to the relevant order $\,n\,$ (determined by the desired order
in $\,\a'\,$) in the ``spacetime fields'' $\,A_a$,\, and then has to integrate
out the world-sheet moduli. This process is medium involved at the technical
level, but the most unsatisfactory aspect of the computation is the absence
of any guiding principle which would allow to ``guess'' the outcome, or at
least to recognise the resulting effective action as ``reasonable''.
\smallskip

The main aim of this paper is to identify such guiding principles. In Sect.\,\ref{sec:sigma},
we will see that the effective action \eqref{SARS} arises
from linearising and quantising a Schild-type world-sheet action for the
WZW model (without adding special boundary terms and without ``detour'' via
CFT), much in the same way as one can obtain the IKKT matrix model \cite{ikkt}
from the $\si$-model for flat strings \cite{yon}. This method is technically
completely straightforward, but remains somewhat mysterious conceptually.

In the sections to follow, we therefore take a closer look at the non-commutativity
of the effective action \eqref{SARS}. In the classical $\si$-model picture,
the symmetry-preserving boundary states mentioned above correspond to branes
whose world-volume is a conjugacy class in the group target \cite{AleksSchom}.
The effective action captures the dynamics of the effective open-string excitations,
gauge bosons and transverse scalars taking values in the (co)tangent and
(co)normal bundles\footnote{Strictly speaking, we are working with the complexified
group, $\,G^\bC$,\, compare the remarks in \cite{ga}.} of the conjugacy class,
respectively. The CFT findings suggest that the open strings ``see'' quantised
world-volumes in the form of ``fuzzy'' conjugacy classes \cite{ars1} -- for
$\,G =SU(2)$,\, these are the well-known fuzzy spheres -- and the action
\eqref{SARS} indeed is a gauge theory on such a non-commutative space --
a rather special gauge theory at that, as we are going to discuss presently.

We will carry out, in Sect.\,\ref{sec:NCGsection}, a systematic reconstruction
of the effective fuzzy geometry of untwisted maximally symmetric D-branes
and the associated gauge dynamics, employing general tools of spectral non-commutative
geometry (NCG). We will show that the effective action derived from CFT is
uniquely determined, among all extensions of the usual Yang--Mills term 
$\,\cS_{\rm YM}[A]\,$ up to order four in the gauge field, by postulating
invariance under an extended gauge symmetry, comprising usual unitary gauge
transformations as well as tangential shifts -- the gauge potential for the
latter is nothing but the Kalb--Ramond field $\,B$.\, This in particular
leads to the fine-tuning between the coefficients of the Yang--Mills term and the Chern--Simons
term in \eqref{SARS}: both are separately invariant under standard gauge
transformations (as first remarked in \cite{klim} for the case of $\,G=SU(2)
\,$) but have mass terms for $\,A\,$ which cancel out (and thus admit moduli
for rigid shifts of branes) only in the particular combination which also
enjoys the extended shift symmetry.

We will point towards the natural classical origin of this extended symmetry
in Sect.\,\ref{sec:gerbe}, which briefly reviews the r\^ole of bundle gerbes
and gerbe modules in WZW models.
\smallskip

A comment is in order concerning the Born--Infeld action, which captures
a great deal of information on how branes, viewed as classical submanifolds
of the target manifold, react to their surrounding target-space geometry.
Applied to Lie groups for targets and conjugacy classes for world-volumes,
the Born--Infeld action allows to understand how the $\,B$-field flux is
capable of balancing the force of gravitational collapse, leading to the
emergence of stable higher-dimensional D-branes on $\,G\,$ \cite{myers,flux-stab,paw-stab,brs-stab}.
One can even ``derive'' the effective action \eqref{SARS} from the Born--Infeld
action by restricting to a stack of D0-branes placed in the background with
non-vanishing Kalb--Ramond field\footnote{The result follows trivially from
the observation that there are no objects charged under the Kalb--Ramond
potential $\,B\,$ in the situation considered and hence the term of the lowest
(linear) order in this field is necessarily proportional to its exterior
derivative, i.e.\ the field strength $\,H$.\, This term augments the standard
(flat-case) Yang--Mills term and thereby reproduces $\,\cS_{\rm ARS}$.}.
We do not, however, regard this as a satisfactory explanation of the specific
form of the low-energy effective action \eqref{SARS} because the argument
relies on inserting D0-branes by hand, while for higher-dimensional WZW branes
the Born--Infeld action gives an intrinsically continuous world-volume geometry
instead of a matrix model as required: It seems that the Born--Infeld approach
is simply ``too classical'' from the outset. 
\medskip

Before we turn to identifying ``reasons'' why \eqref{SARSX} or \eqref{SARS}
is in fact a very natural effective action, let us introduce some of its
further characteristics. Its equations of motion can be written as
\qq\label{Fuzzeq}
\bigl[\,X_a,[\,X_a,X_b\,]-\sfi\,f_{abc}\,X_c\,\bigr]=0\,,\qquad{\rm or} 
\qquad[\,Y_a+A_a,F_{ab}(A)\,]=0\,; 
\qqq
the second form simply means that $\,F\,$ is covariantly constant. These
equations were examined in great detail in \cite{Fred} (but see also \cite{StefanVolker,Antonetal,Monnier}),
where it was in particular shown that solutions of the form
\qq\label{Fsym}
F_{ab}(A)=\Phi_{ab}\cdot\bd1_N\,,\qquad\Phi_{ab}=-\Phi_{ba}\in\bC
\qqq
precisely correspond to maximally symmetric WZW branes, labelled by a positive
weight $\,\la\in P_+(\ggt)\,$ and associated, \`a la Kirillov, to the conjugacy
class $\,\cC_\la\subset G\,$ wrapped by the (classical) D-brane. The correspondence
to single conjugacy classes holds as long as $\,N=n\cdot\dim V_\la\,$ is
a multiple of the dimension of an irreducible $\,\ggt$-module, where $\,n\,$
is the number of Chan--Paton labels in a stack of D-branes wrapping the same
conjugacy class $\,\cC_\la$.\, Using techniques established for the boundary
CFT description of the Kondo effect \cite{AfflLud}, it was argued in \cite{ars2,Fred,StefanVolker}
that one can reach superpositions of stacks over different conjugacy classes
upon brane condensations, i.e.\ renormalisation-group flows. These condensations
can in particular change the dimension of the conjugacy class: higher-dimensional
WZW branes can be viewed as bound states of D0-branes. 

The relation between \eqref{Fsym} and maximally symmetric branes was established
in \cite{ars2,Fred} by comparing the value of the effective action at such
a configuration to the known $\,g$-factors (or boundary entropies) for Cardy
boundary states. Note that the equations of motion \eqref{Fuzzeq} allow for
other solutions, but since the action \eqref{SARS} was derived in conformal
perturbation theory around a maximally symmetric boundary condition, it will
only see perturbative renormalisation-group fixed points.

Any gauge field configuration can be split into a traceless and a trace part,
\qq\label{split}
A_a=A^0_a+A^T_a\,,\qquad{\rm with}\quad A^T_a:=\frac{1}{N}\,\tr\,A_a\cdot
\bd1_N\,.
\qqq
Using the Bianchi identity for $\,F_{ab}$,\, it can be shown \cite{Fred}
that whenever $\,A_a\,$ solves the equations of motion \eqref{Fuzzeq} then
$\,F_{ab}(A^0)=0$.\, If $\,A_a\,$ moreover satisfies the maximal symmetry
condition \eqref{Fsym} then $\,\Phi_{ab}\equiv f_{abc}\,A^T_c$.

What is more relevant for our purposes is to note that the split \eqref{split}
is preserved by the following {\sl extended gauge symmetry} of $\,\cS_{\rm
ARS}$,\, parametrised by standard gauge transformations $\,U\in SU(N)\,$
{\sl and} scalar shifts given by $\,\Pi_a\in\bC$:
\qq\nn
A_a\too A_a^{U,\Pi}:=U^\dagger\,A_a\,U+\sfi\,U^\dagger\,[\,Y_a,U\,]+\Pi_a
\cdot\bd1_N\,.
\qqq
Indeed, separating $\,A_a^{U,\Pi}\,$ into its traceless and trace parts,
we obtain
\qq\nn
A_a^0\too U^\dagger\,A_a^0\,U+\sfi\,U^\dagger\,[\,Y_a,U\,]\,,\non\non
A_a^T\too A_a^T+\Pi_a\cdot\bd1_N\,,\nonumber
\qqq
which identifies $\,A^0\,$ as an $\,SU(N)\,$ gauge field. The fields $\,A^T_a$,\,
on the other hand, are scalars under the gauge symmetry $\,SU(N)\,$ and can
be viewed, physically, as the translational moduli associated with rigid
one-sided $\,G$-translations in the target. By selecting some maximally symmetric
D-brane (i.e.\ some conjugacy class), this global symmetry is broken so that
 $\,A^T\,$ becomes a Goldstone vector boson.

\section{Fuzzy matrix dynamics directly from the WZW \boldmath $\si$-model}\label{sec:sigma}

\noindent
In this section, we establish a technically straightforward but nevertheless
conceptually surprising relation between the microscopic description of closed-string
propagation in the Lie-group target and the fuzzy matrix model \eqref{SARS}.
We start from the WZW world-sheet action for closed strings, linearise around
constant maps, then pass to a classically equivalent action of the Schild
type. Next, one can subject the dynamical variables of the latter to a natural
quantisation procedure, involving a truncation to finitely many degrees of
freedom on the way. In this manner, one obtains the effective action \eqref{SARS}
for open strings attached to maximally symmetric branes in the WZW target.
Boundary conditions (which were originally left unspecified) arise naturally
as a consistency condition in the process of truncation quantisation.

This procedure is a generalisation of Yoneya's derivation of the IKKT matrix
model action (the D-instanton analogue of the BFSS matrix model) from the
classical world-sheet action \cite{yon}. Our derivation in particular allows
to see the connections between the symmetries of the WZW model and torsion
geometry very clearly, as will be described in Sect.\,\ref{sub:torgeo}.

\subsection{From the linearised WZW model to the Schild action}

\noindent
We start on a closed Euclidean world-sheet $\,\Si\,$ with coordinates   
$\,\si=(\si_1,\si_2)\,$ and metric $\,\g=\gamma_{AB}\,\sfd\si^A\,\sfd\si^B$.\,
The action for the level-$\sfk\,$ WZW model is
\qq
\cS_{\rm WZW}[g,\gamma]=-\frac{\sfk}{4\pi\a'}\,\int_\Si\,\sfd^2\si\,    
\sqrt{\det\g}\,(\g^{-1})^{AB}\,\tr\,\left[\lb g^{-1}\,\p_A g\rb\,\lb g^{-
1}\,\p_B g\rb\right]+\frac{\sfi\sfk}{4\pi\a'}\,\int_\Si\,g^*\sfd^{-1}\chi
(g)\,,\cr\label{bWZW}
\qqq
with
\qq\nn
\chi(g)=\frac{1}{3}\,\tr\,\left[\lb g^{-1}\,\sfd g\rb\wedge\lb g^{-1}\, 
\sfd g\rb\wedge\lb g^{-1}\,\sfd g\rb\right]\,,\qquad g\in G\,.
\qqq
We expand the maps $\,g(\si)\,$ around some fixed group element $\,g_0\in
G$,\, i.e.\ consider
\qq\label{expl}
g(\si)=:g_0\cdot\ee^{\sfi\,\th^a\,\widetilde X_a(\si)}=g_0\cdot\left[e+\sfi\,
\th^a\,\widetilde X_a(\si)+\cO\lb\ell^{-2}\rb\right]\,,
\qqq
where $\,e\in G\,$ is the group unit, $\,\th^a\,$ form an orthonormal basis
of the tangent space of $\,G\,$ at $\,g_0\,$ (a one-sided translate of $\,T_e
G\cong\ggt\,$) and satisfy the standard structure relations
\qq\nn
[\,\th^a,\th^b\,]=\sfi\,f_{abc}\,\th^c\,;
\qqq
the objects 
\qq\nn
\widetilde X_a\equiv\frac{1}{\ell}\cdot X_a\,,
\qqq
are dimension-less world-sheet embedding fields. The appearance of the stringy
length scale $\,\ell=\sqrt{\a'\,\sfk}\,$ makes it clear that the linearisation
of the WZW action is in fact a long-distance or low-energy approximation.

Substituting the expansion \eqref{expl} into \eqref{bWZW} and keeping only
terms up to the order $\,\ell^{-3}$,\, we obtain the linearised world-sheet
model\footnote{The contribution of the kinetic term in $\,\cS_{\rm WZW}\,$
trilinear in $\,X\,$ turns out to be traceless.}
\qq
\cS_{\rm lin}[X;\g]=\cN\,\int_\Si\,\sfd^2\si\,\lb\sqrt{\det\g}\,(\g^{-1}
)^{AB}\,\d_{ab}\,\p_A X_a\,\p_B X_b+\frac{\sfi}{3\ell}\,\ep^{AB}\,f_{abc}
\,X_a\,\p_A X_b\,\p_B X_c\rb\cr\label{Slinf}
\qqq
with a normalisation constant $\,\cN\equiv\frac{1}{4\pi\a'{}^2}$.

In the next step, we apply arguments from \cite{yon} and establish classical
equivalence between $\,\cS_{\rm lin}\,$ and the Schild-type action \cite{schild}
\qq
\cS_{\rm Schild}[X;e]=\int_\Si\,\sfd^2\si\,e\lb\frac{1}{4}\,\d_{ac}\,
\d_{bd}\,\{X_a,X_b\}_e\,\{X_c,X_d\}_e+\frac{\sfi}{3\ell}\,f_{abc}\,X_a\{
X_b,X_c\}_e+\mu\rb\,,\cr\label{Schild}
\qqq
written in terms of an auxiliary field $\,e=e(\si)\,$ -- which is a positive
scalar density of weight $\,w(e)=-1\,$ on the world-sheet --, a positive constant
$\,\mu>0\,$ and the so-called Nambu--Poisson bracket
\qq\label{NPBe}
\{f,g\}_e:=\frac{1}{e}\,\ep^{AB}\,\p_A f\,\p_B g
\qqq
on the algebra of smooth functions on the world-sheet. The proof of the equivalence
between $\,\cS_{\rm lin}\,$ and $\,\cS_{\rm Schild}\,$ proceeds by introducing
a third, intermediate model
\qq
\cS_\sim[X;e,t]=\int_\Si\,\sfd^2\si\,\frac{1}{e}\,\lb-\det t+\frac{t^{AB}
}{\sqrt{2}}\,\d_{ab}\,\p_A X_a\,\p_B X_b+\frac{\sfi e^2}{3\ell}\,f_{abc}\,
X_a\,\{X_b,X_c\}_e+e^2\,\mu\rb\cr\label{Sinter}
\qqq
with $\,e\,$ and $\,\mu\,$ as above and $\,t^{AB} = t^{BA}\,$ another auxiliary
field (a tensor density of weight $\,w(t)=-2\,$ on the world-sheet).
Passing to
\qq\nn
\widetilde t^{AB}:=t^{AB}-\frac{1}{\sqrt{2}}\,\ep^{AC}\,\ep^{BD}\,\d_{ab}\, 
\p_C X_a\,\p_D X_b\,,
\qqq
we readily check the identity
\qq\nn
\cS_\sim[X;e,t]\equiv\cS_{\rm Schild}[X;e]-\int_\Si\,\sfd^2\si\,\frac{\det
\widetilde t}{e}\,,
\qqq
and (classical) equivalence of these two models follows\footnote{Strictly
speaking, we have only proved the existence of a bijection between classical
configurations of the two models. The latter compose the respective phase
spaces which are endowed with additional structure -- the (pre)symplectic
structure, central to the quantisation of the models. The mapping discussed
can actually be demonstrated to define a symplectomorphism between the two
phase spaces, and it provides an example of a completely general symplectomorphic
equivalence between the Polyakov and the Schild formulation valid for an
arbitrary two-dimensional (dilaton-free) non-linear $\si$-model, discussed
in  \cite{rSr2007tap}.}
since the equation of motion for $\,\widetilde t^{AB}\,$ is simply $\,\widetilde
t^{AB}=0$.

Next, we rescale the dynamical fields as $\,X\too\sqrt{\frac{\mu}{2}}\cdot
X\,$ and perform a change of variables $\,(e,t)\too(\widetilde e,\g)\,$ in
\eqref{Sinter}, with $\,\widetilde e\,$ a world-sheet scalar and $\,\g=(\g_{AB})\,$
a metric tensor on $\,\Si$,\, such that
\qq\nn
e=e(\widetilde e,\g)=\widetilde e\,\sqrt{\det\g}\,,\qquad t=t(\widetilde
e,\g)=\lb\widetilde e^2\,\det\g\rb\cdot\g^{-1}\,.
\qqq
In this way, we obtain
\qq
\cS_\sim[X;e,t]&=&\lb\frac{\mu}{2}\rb^{\frac{3}{2}}\,\int_\Si\,\sfd^2\si\,
\sqrt{\det\g}\,\lb\frac{\widetilde e}{\sqrt{\mu}}\,(\g^{-1})^{AB}\,\d_{ab}\,
\p_A X_a\,\p_B X_b+\frac{\sfi}{3\ell}\,f_{abc}\,X_a\,\{X_b,X_c\}_\g\rb\cr\cr
&\phantom{-}&\quad\quad-\int_\Si\,\sfd^2\si\,\sqrt{\det\g}\,\widetilde e\,\lb
\widetilde e^2-\mu\rb\,,\label{Stil}
\qqq
where now the Nambu-Poisson bracket is defined using $\,\g$,
\qq\nn
\{f,g\}_\g=\frac{1}{\sqrt{\det\g}}\,\ep^{AB}\,\p_A f\,\p_B g\,.
\qqq
Solving the equation of motion for the trace part of $\,\g$,
\qq\nn
(\g^{-1})^{AB}\,\frac{\d\cS_\sim}{\d(\g^{-1})^{AB}}\must 0\,,
\qqq
we obtain
\qq\nn
\widetilde e=\sqrt{\mu}
\qqq
by virtue of the assumed positivity of $\,\widetilde e$.\, Plugging this result
back into \eqref{Stil}, we arrive at the classically equivalent model
\qq\nn
\bar\cS_\sim[X;\g]=\lb\frac{\mu}{2}\rb^{\frac{3}{2}}\,\int_\Si\,\sfd^2\si
\,\sqrt{\det\g}\,\lb(\g^{-1})^{AB}\,\d_{ab}\,\p_A X_a\,\p_B X_b+
\frac{\sfi}{3\ell}\,f_{abc}\,X_a\,\{X_b,X_c\}_\g\rb\,,\cr
\qqq
which reproduces -- up to an irrelevant rescaling -- the action \eqref{Slinf},
\qq\nn
\bar\cS_\sim[X;\g]\equiv\frac{1}{\cN}\lb\,\frac{\mu}{2}\rb^{\frac{3}{2}}\,
\cS_{\rm lin}[X;\g]\,.
\qqq
Thus, we have established the classical equivalence of the linearised WZW action
and the Schild action,
\qq\nn
\cS_{\rm Schild}[X;e]\quad\sim\quad\cS_{\rm lin}[X;\g]\,,
\qqq
which leaves us the Schild model \eqref{Schild} to play with further.

\subsection{A matrix model for D0-branes}

\noindent
The crucial property of the Schild model is that it is formulated entirely
in terms of the embedding field $\,X\,$ and its world-sheet Nambu--Poisson
brackets \eqref{NPBe}. The latter admit a straightforward canonical quantisation
\qq\label{canq}
\{\cdot,\cdot\}_e\too-\sfi\,[\,\cdot,\cdot\,]\,.
\qqq
In addition, we make the replacement \cite{hoppe}
\qq\label{inTr}
\int_\Si\,\sfd^2\si\,e\too\Tr_\ceH
\qqq
of the world-sheet integral by a trace over the ensuing Hilbert space $\,\ceH\,$
-- which is the standard replacement accompanying canonical quantisation:
the integral \wrt the Liouville measure over the classical ``phase space''
to be quantised (here, it is simply the world-sheet $\,\Si\,$ equipped with
the Nambu-Poisson structure) becomes the trace over the Hilbert space. So
from the formal point of view, the transition \eqref{canq} and \eqref{inTr}
is tantamount to the canonical quantisation of the simple symplectic structure
on $\,\Si\,$ associated with the Nambu-Poisson bracket \eqref{NPBe} on the
algebra $\,C^\infty(\Si)\,$ of smooth functions on the world-sheet.

We can now furthermore choose a ``regularisation'' by demanding that $\,\ceH\,$
is in fact finite-dimensional -- which is a reasonable choice (and the typical
outcome) for a compact phase space. This makes the $\,X\,$ into matrices
and takes us from the original world-sheet model to a matrix model, namely
\qq\nn
\cS_M[X]=\Tr_\ceH\,\lb-\frac{1}{4}\,\d_{ac}\,\d_{bd}\,[\,X_a,X_b\,]\,[\,
X_c,X_d\,]+\frac{1}{3\ell}\,f_{abc}\,X_a\,[\,X_b,X_c\,]+\mu\rb\,,
\qqq
which is, up to a rescaling $\,(X_a,\mu)\too (-\sfi\,\ell^{-1}\,X_a,\ell^{-4}
\,\mu)\,$ and a trivial overall normalisation, nothing but the fuzzy matrix
model \eqref{SARSX},
\qq\nn
\cS_{\rm ARS}[X]\equiv\ell^4\,\cS_M[X]\,.
\qqq

\smallskip

Let us now assume that the world-sheet $\,\Si\,$ has a boundary and define
the original WZW action \eqref{bWZW} there, without adding any boundary terms
(which should, of course, be done to obtain a well-defined path integral
formulation, see the brief review in Sect.\,\ref{sec:gerbe} below). In this setting, the ``regularisation'' chosen above, which makes the
$X$ into bounded operators, has some immediate consequences. The basic property of the trace
\qq\nn
\Tr_\ceH\,\lb[\,X_a,X_b\,]\rb=0
\qqq
translates back into an integral of the Poisson bracket, leading to 
the condition \cite{solo}
\qq\label{Solov}
\int_{\p\Si}\,\sfd t\,X_b\,\p_t X_a=0
\qqq
on the variables in the Schild action. 
Thus, one is led to Dirichlet conditions in all directions
\qq\nn
\p_t X_a\big\vert_{\p\Si}=0\,,
\qqq 
as the natural (and probably the only admissible) choice of boundary conditions over the world-sheet boundary. 

Note the difference between \eqref{Solov} and the boundary term 
of the variation in, say, a free boson theory on the upper half 
plane:\ the latter allows for both Dirichlet and Neumann boundary  
conditions. Note also that Dirichlet conditions in all directions 
are the ones that 
solve the boundary variation problem in any non-linear $\si$-model, 
while Neumann conditions are generically obstructed by the $B$-field. 

Our little derivation leading from the world-sheet action of the WZW model to the open-string effective action in addition
suggest that the latter is a model describing D0-brane dynamics -- completely
in line with the CFT result that higher-dimensional branes can be obtained
as bound states of D0-branes.

\subsection{The torsion geometry of the linearised \boldmath $\si$-model}\label{sub:torgeo}

\noindent
So far, we have focused on the canonical and algebraic structure of the models
involved, leaving aside the geometric interpretation of their building blocks
-- in particular of the metric and the $\,B$-field. While most of the following
remarks are true in arbitrary conformally invariant $\si$-models, the linearised
WZW action \eqref{Slinf} offers a convenient starting point to discuss symmetries
and to illuminate the geometric r\^ole of the Kalb--Ramond field $\,B\,$
from this perspective. The wider context of the ensuing interpretation is
given by the ``geometrostasis'' ideas from \cite{bracurzach}. For us, its
main importance lies in the fact that it naturally leads to the notion of
a torsion gerbe (present in all $\si$-models with a curved target, in particular
WZW models), and lifting the differential-geometric structure of a bundle
gerbe on the group manifold to the fuzzy r\'egime will eventually shed new
light on the effective open-string dynamics of WZW D-branes.

To get there, we first study symmetries of the model, as descended from those
of its complete (group-integrated) version \eqref{bWZW}. It is easy to check
that the symmetries which survive the linearisation are the tangent-space
counterparts of the left-right shifts by group elements constant on the world-sheet.
Thus, the left-right affine-algebra symmetry $\,\kmg^L\ox\kmg^R\,$ of the
WZW model reduces to its horizontal component, and its action splits into
a vector (adjoint) part %$(\ggt^L \ox \ggt^R)^V$,
\qq\nn
X_a\,t_a\equiv X\too X+\sfi\,[\,\La,X\,]=(X_a+f_{abc}\,X_b\,\La_c)\,t_a\,,
\qquad{\rm with}\quad\La\equiv\La_a\,t_a\in\ggt\,,
\qqq
%which can be integrated to the adjoint group action  $(G^L \x G^R)^V$,
%\beq
%X \longmapsto U X U^{-1}, \qquad \qquad U \in G
%\eeq
and the one-sided $\,\ggt$-shifts
\qq\nn
X\too X+\Pi\,,\qquad{\rm with}\quad\Pi\equiv\pi_a\,t_a\in\ggt
\qqq
which come from one-sided group translations (neither $\,\La\,$ nor $\,\Pi\,$
depend on the world-sheet variable $\,\si\,$). It is important to note that
both symmetries are present in an unaltered form in the Schild action \eqref{Schild}
as well.

For WZW models, the Cartan--Killing metric and the Kalb--Ramond field
locally take the form \cite{cornalba,Bordalo:2004xg} 
\qq\label{Binvel}
\tx g_{ab}=\d_{ab}+\cO(\ell^{-2})\,,\qquad\qquad B_{ab}=\frac{1}{3\ell}\,f_{a
bc}\,X_c+\cO(\ell^{-3})\,,
\qqq
and to derive their transformation properties, we treat them as (tangent-space)
tensors and so simply compute their Lie derivatives in the direction of the
relevant variation vector fields,
\qq\nn
\d^{\rm iso}_\La X_a=f_{abc}\,X_b\,\La_c\,,\qquad\qquad\d^{\rm axi}_\Pi X_a
=\pi_a\,.
\qqq
(The notations ``iso'' and ``axi'' are borrowed from  \cite{bracurzach} and
stand for the adjoint and shift symmetry, respectively.) In this way, we
obtain
\qq\nn
\d^{\rm iso}_\La\tx g_{ab}=0=\d^{\rm axi}_\Pi\tx g_{ab}\,,\qquad\qquad\d^{\rm 
iso}_\La B_{ab}=0\,,
\qqq
and
\qq\label{compareB}
\d^{\rm axi}_\Pi B_{ab}=-\p_{[a}\widetilde\pi_{b]}\,,\qquad{\rm with}\quad  
\widetilde\pi_a=-\frac{1}{3\ell}\,f_{abc}\,X_b\,\pi_c\,.
\qqq
The lesson to draw from this simple calculation is that the torsion potential
$\,B\,$ is a gauge field for tangential translations, under which it transforms
simply as
\qq\label{Bastor}
B\xrightarrow{\Pi}B-\sfd\widetilde\Pi\,.
\qqq
This is, indeed, the fundamental r\^ole of the torsion potential in gauge
approaches to gravity, and - in particular - in the teleparallel gravity
theory for so-called Weitzenb\"ock geometries, which work with flat connections
with torsion instead of (but equivalent to) the more familiar Levi-Civita
connections. The latter fits perfectly with the effective equations of motion
in string theory derived from the vanishing of the beta function (to first order in $\,\a'\,$), which require (to have non-anomalous conformal symmetry)
that the curvature is precisely cancelled by the torsion field $\,H$,\, for
which the Kalb--Ramond field is a local potential; equivalently, one can
augment the standard Levi-Civita connection of the metric $\,\tx g\,$ by the
torsion field in such a way as to form a (Weitzenb\"ock) connection with
a vanishing Riemann tensor.  It is this phenomenon of cancellation between
the two contributions to the Riemann curvature, the (metric) Levi-Civita
one and the (topological) torsion one, required by non-anomalous conformal
symmetry of the underlying $\si$-model, which was termed ``geometrostasis''
in the original papers\footnote{For further literature on teleparallel gravity,
see \cite{curzach,agip,telintro}. Gauge theories of gravity were recently
reviewed in \cite{hehl}.} \cite{bracurzach}.

\section{WZW models and bundle gerbes}\label{sec:gerbe}

\noindent
The gauge freedom intrinsic to the definition of the torsion field above,
and even the sheer presence of torsion, find their natural origin in the
gerbe structure underlying the topological WZ term in the WZW action functional.
Whenever one considers string targets with a non-trivial $\,B$-field, issues
like gauge invariance require a more careful analysis, taking into account
extra global structures in the form of a bundle gerbe \cite{gawtop,gawreis,gawAb}.
In this section, we briefly review the necessity to consider gerbes in string
theory, mainly following \cite{gawtop,gawreis,gawAb} and \cite{alva}. We
also recall the defining properties of a bundle gerbe $\,\xcG_H$:\, In the
context of non-linear $\si$-models, bundle gerbes are the natural differential-geometric
structure induced in the target by the presence of a closed non-trivial torsion
3-form $\,H\,$ (for which $\,B\,$ is a local potential). In direct analogy
with line bundles, bundle gerbes comprise (Deligne) cohomology classes and
the relevant cohomological equivalence (the so-called stable isomorphism)
manifests itself in particular through the translational gauge symmetry
discussed in the previous section. We will implant this bundle gerbe gauge
symmetry, in Sect.\,\ref{sec:NCGsection}, into the non-commutative framework
and arrive at a symmetry principle that singles out the low-energy effective
action of open-string theory within NCG.

\subsection{A lightning course on bundle gerbes}

\noindent
Let us start by recalling a few basic facts about bundle gerbes using their
local description, as detailed, e.g., in \cite{jlB}. Thus, a bundle gerbe
$\,\xcG_H\,$ associated to the closed 3-form background $\,H\,$ (the globally
defined curvature of $\,\xcG_H\,$) is -- from the physical point of view
-- a construct\footnote{Bundle gerbes were introduced by Giraud in \cite{giraud}
and later rephrased and developed by Brylinski in \cite{jlB}, where contact
was made with the cohomological approach to WZW models due to Alvarez \cite{alva}
and Gaw\c{e}dzki \cite{gawtop}. Strictly speaking, their application in the
WZW setting requires a reformulation of the original concept in the more
natural geometric language of Murray \cite{murray,murrays}, see also \cite{murson}
for an introduction to the subject, but we shall not need the complete picture
in what follows.} which enables us to give meaning to topological actions,
of the kind of the WZ term in \eqref{bWZW}, in topologically nontrivial geometries
$\,\cM$.\, An example of such a geometry is a compact Lie group. Here, $\,\cM\,$
is the target space of the $\si$-model, in which the world-sheet $\,\Si\,$
is embedded {via}
\qq\nn
\phi\ :\ \Si\too\cM\,.
\qqq
The obvious problem that we encounter when trying to give meaning to the
WZ term is the lack of a global potential for $\,H$.\, The solution to this
problem valid in all generality\footnote{A solution applicable in the case
of a simply connected Lie-group target boils down to the introduction of
a filling manifold for $\,\Si$,\, together with an appropriate extension
of $\,\phi$,\, and was proposed already in the original paper \cite{wzwitt}.
This prescription fails in the non-simply connected case.} uses \emph{local
data} for the bundle gerbe, as captured by the Deligne cohomology group $\,H^2(\cM,\cD^{(2)})\,$
based on the differential complex
\qq\nn
\cD^{(2)}\ : \ \unl{U(1)}\xrightarrow{\frac{1}{\sfi}\sfd\log}\unl{\Om}^1(
\cM)\xrightarrow{\ \ \sfd\ \ }\unl{\Om}^2(\cM)\,.
\qqq
Above, $\,\unl{U(1)},\,\unl{\Om}^1(\cM)\,$ and $\,\unl{\Om}^2(\cM)\,$ are
the sheaves of smooth $\,U(1)$-valued functions, smooth 1-forms and smooth
2-forms on $\,\cM$,\, respectively. The local data are given by a family
of triples $\,(B_i,\a_{ij},g_{ijk})_{i,j,k\in\cI}\in\vC^0(\xcU,\unl{\Om}^2)
\oplus\vC^1(\xcU,\unl{\Om}^1)\oplus\vC^2(\xcU,\unl{U(1)})\,$ of elements
of the sets of \Cv ech $\,p$-cochains $\,\vC^p(\xcU,A)\,$ with values in
the appropriate sheaves $\,A\in\{\unl{U(1)},\unl{\Om}^1,\unl{\Om}^2\}$.\,
The \Cv ech cochains are defined in the standard way with reference to a
good covering $\,\xcU\equiv\{\xcU_i\}_{i\in\cI}\,$ of $\,\cM$,\, that is
an open covering such that all non-empty multiple intersections $\,\xcU_{i_1
i_2\ldots i_n}\equiv\xcU_{i_1}\cap\xcU_{i_2}\cap\ldots\cap\xcU_{i_n}\neq
\emptyset\,$ are contractible. A multi-indexed object $\,X_{i_1 i_2\ldots
i_n}\,$ is defined on the non-empty multiple overlap $\,\xcU_{i_1 i_2\ldots
i_n}\,$ and satisfies $\,X_{i_{\si(1)}i_{\si(2)}\ldots i_{\si(n)}}=\sign(
\si)\,X_{i_1 i_2\ldots i_n}\,$ for any permutation $\,\si\in\gt{S}_n\,$ of
the indices.

In terms of the local data, the gerbe can be viewed as a collection of so-called
curvings $\,B_i$,\, connections $\,\a_{ij}\,$ and transition functions $\,g_{ijk}$,\,
defined patch-wise and related by the constraints
\qq\nn
\left\{\barr{l}
H\vert_{\xcU_i}=:\sfd B_i\\ \\
(B_j-B_i)\vert_{\xcU_{ij}}=:\sfd\a_{ij}\\ \\
(\a_{jk}-\a_{ik}+\a_{ij})\vert_{\xcU_{ijk}}=:\sfi\,\sfd\log g_{ijk}
\\ \\
(g_{jkl}\,g^{-1}_{ikl}\,g_{ijl}\,g^{-1}_{ijk})\vert_{\xcU_{ijkl}}\must 1
\earr\right.
\qqq
modulo the equivalences
\qq\label{bgDeq}
\left\{\barr{l}
B_i\longrightarrow B_i-\sfd\widetilde\Pi_i\\ \\
\a_{ij}\longrightarrow(\a_{ij}-\widetilde\Pi_j+\widetilde\Pi_i+\sfi\,\sfd\log\widetilde
\chi_{ij})\vert_{\xcU_{ij}}\\ \\
g_{ijk}\longrightarrow(g_{ijk}\,\widetilde\chi_{jk}\,\widetilde\chi^{-1}_{ik}\,\widetilde
\chi_{ij})\vert_{\xcU_{ijk}}
\earr\right.
\qqq
valid for any $\,(\widetilde\Pi_i,\widetilde\chi_{ij})_{i,j\in\cI}\in\vC^0(\xcU,
\unl{\Om}^1)\oplus\vC^1(\xcU,\unl{U(1)})$.\, The main point of the local
differential-geometric structure on $\,\cM\,$ thus introduced is that it
provides an unambiguous definition (i.e. it assigns a $\,c$-number value)
to the topological WZ amplitude
\qq\label{bulkhol}
\ee^{\sfi\,\int_\Si\,\phi^*\sfd^{-1}H}\equiv\prod_{t\in\triangle_\Si}\, 
\ee^{\sfi\,\int_t\,\phi^*_t B_{i_t}}\,\prod_{e\subset t}\,\ee^{\sfi\,   
\int_e\,\phi^*_e\a_{i_t i_e}}\,\prod_{v\subset e\subset t}\,g^{\ep(v    
)}_{i_t i_e i_v}(\phi(v))\equiv\Hol_{\cG_H}(\phi)\,,
\qqq
independent of all choices made \emph{as long as the world-sheet is closed},
$\,\p\Si=\emptyset$.\, Here, $\,\triangle_\Si=\{t,e,v\}\,$ is a triangulation
of $\,\Si\,$ compatible with $\,\xcU\,$ in the sense that the plaquettes
$\,t$,\, their edges $\,e\,$ and vertices $\,v\,$ are chosen such that for
each of them, $\,f\in\triangle_\Si$,\, we have $\,\phi(f)\subset\xcU_{i_f}\,$
for some element of the covering. Whenever the orientation of $\,v$,\, as
inherited from $\,t\,$ via $\,e$,\, is negative, we have $\,\ep(v)=-1$,\,
otherwise $\,\ep(v)=1$.\, Finally, $\,\phi_f\equiv\phi\vert_f\,$ are the
respective restrictions of the embedding map. The above form of the WZ amplitude
defines the holonomy $\,\Hol_{\cG_H}(\phi)\,$ of the bundle gerbe $\,\cG_H\,$
over the world-sheet. The fundamental r\^ole of the amplitude in the (quantised)
theory follows from the fact that it enters the path-integral definition
of correlations functions of local operators $\,\widehat\cO_i\,$ as 
\qq\label{fullamp}
\big\lan\prod_I\,\widehat\cO_I\big\ran\equiv\int\,\xcD\phi\,\ee^{-\cS_{kin}[\phi]}
\cdot\ee^{-\sfi\,\int_\Si\,\phi^*\sfd^{-1}H}\cdot\prod_I\,\cO_I(\phi)\,.
\qqq

\subsection{Twisted gauge fields on \boldmath $\,\cG_H$-branes}\label{sub:hola}

\noindent
Now that we have acquainted ourselves with the closed-string case, let us
consider a boundary CFT describing open strings, defined on a world-sheet
whose  boundary $\,\p\Si=\bigsqcup_\a\,S_\a\neq\emptyset\,$ may have several
connected components $\,S_\a$.\, In this setting, the requirement of the
unambiguity
of the WZ amplitude necessitates the introduction of extra geometric structure
-- a set of \emph{gerbe modules} $\,\cE_\a\too\xcD_\a$,\, also termed twisted
gauge bundles \cite{kapu}, one for each of the \emph{$\,\cG_H$-brane} submanifolds
$\,\xcD_\a\supset\phi(S_\a)\,$ in which the corresponding boundary components
$\,S_\a\,$ are embedded by $\,\phi\vert_{S_\a}\equiv\phi_\a$.\, A correction
to the bulk expression \eqref{bulkhol} for the WZ amplitude given by the
product of their holonomies along the respective boundary components of the
world-sheet renders the amplitude unambiguous by removing its dependence
on the choice of local data for the bundle gerbe $\,\cG_H\,$ in the presence
of the world-sheet boundary. The latter property of the correction follows
straightforwardly from the twisted character of the gluing rules
\qq
(\tx A_j-\tx G_{ij}^{-1}\,\tx A_i\,\tx G_{ij}-\sfi\,\tx G_{ij}^{-1}\,\sfd
\tx G_{ij}+\a_{ij}\ox\bd1_N)\big\vert_{\xcO_{ij}}=0\,,\non\label{Atw}\\ 
(\tx G_{jk}\,\tx G_{ik}^{-1}\,\tx G_{ij}-g_{ijk}\ox\bd1_N)\big\vert_{\xcO_{ijk}}
=0\nonumber
\qqq
for the (local) gauge fields $\,\tx A_i\,$ of the twisted gauge bundles.
Here, these fields have been defined in terms of the local data for (a specific)
$\,\cE_\a$.\, The data consist of a family of pairs $\,(\tx A_i,\tx G_{ij})
\in\vC^0(\xcO,\unl{\Om}^1\ox\gt{u}(N))\oplus\vC^1(\xcO,\unl{U(N)})\,$ associated
to a good covering $\,\xcO\equiv\{\xcO_i\}_{i\in\cI_\a}\,$ of the submanifold
$\,\xcD_\a\,$ (compatible, in the obvious sense, with $\,\xcU\,$).

Given a suitable triangulation $\,\triangle_{S_\a}=\{e_\a,v_\a\}\,$ of the
connected component $\,S_\a\,$ of $\,\p\Si=\bigsqcup_\a\,S_\a$,\, with $\,e_\a\,$
denoting boundary edges and $\,v_\a\,$ the boundary vertices, the holonomy
of the corresponding gerbe module $\,\cE_\a\,$ over $\,S_\a\,$ can be expressed
succinctly as 
\qq\nn
\Hol_{\cE_\a}(\phi_\a)=\tr_{U(N)}\,\lb P\prod_{e_\a\in\triangle_{S_\a}}\,
\ee^{\sfi\,\int_{e_\a}\,\phi^*_{e_\a}\tx{A}_{i_{e_\a}}}\,\prod_{v_\a\subset
e_\a}\,\tx{G}^{\ep(v_\a)}_{i_{v_\a}i_{e_\a}}(\phi_\a(v))\rb\,,
\qqq
where $\,P\,$ denotes the standard path ordering. As previously, $\,\ep(
v_\a)=-1\,$ for a negatively oriented vertex and $\,\ep(v_\a)=1\,$ otherwise.
The above holonomy completes the definition of the WZ amplitude in the boundary
case as per
\qq\nn
\ee^{\sfi\,\int_\Si\,\phi^*\sfd^{-1}H}\equiv\Hol_{\cG_H}(\phi)\cdot\prod_\a
\Hol_{\cE_\a}(\phi_\a)\,,
\qqq
which is now free of ambiguities.
\smallskip

The central distinguishing feature of the twisted gauge field is its dependence
on the gerbe data, as reflected in the transformation rules \eqref{Atw}.
The torsion potential is seen to exchange, via the $\,\a_{ij}$-terms, some
of its degrees of freedom with the gauge field. This leads us to consider
non-trivial transformations of $\,\tx A_i\,$ under tangent-space translations,
\qq\label{APik}
\tx A_i\too\tx A_i+\widetilde\Pi_i\ox\bd1_N\,,
\qqq
with $\,\widetilde\Pi_i\,$ defining, as previously, the gauge shift of the torsion
potential (curving) $\,B_i$.\, In this way, we uncover a scalar-shift \emph{extension
of the standard gauge symmetry} as a direct consequence of the twisted nature
of the gauge fields supported by $\,\cG_H$-branes. The presence of this extension
compels us to consider two possible definitions of the field-strength tensor
- the ``standard'' one, $\,\tx F_i=\sfd\tx A_i+\sfi\,\tx A_i^2$,\, with twisted
gluing/gauge-transformation properties 
\qq\label{gaugecltw}
(\tx F_j-\tx G_{ij}^{-1}\,\tx F_i\,\tx G_{ij}+\sfd\a_{ij}\ox\bd1_N)     
\big\vert_{\xcO_{ij}}=0\,,\qquad\qquad\tx F_i\too\tx H_i^{-1}\,\tx F_i\,
\tx H_i+\sfd\widetilde\Pi_i\ox\bd1_N\,,
\qqq
alongside a twisted one, $\,\cF_i=\tx F_i+B_i\ox\bd1_N$,\, with the ``standard''
gluing/gauge-transformation properties
\qq\label{Ftw}
(\cF_j-\tx G_{ij}^{-1}\,\cF_i\,\tx G_{ij})\big\vert_{\xcO_{ij}}=0\,,\qquad
\qquad\cF_i\too\tx H_i^{-1}\,\cF_i\,\tx H_i\,.\nonumber
\qqq
(above, $\,\tx H_i\in \unl{U(N)}\,$ define the standard (unitary) component
of a gauge transformation). The latter is the familiar combination of the
gauge field and the Kalb--Ramond
background appearing, in particular, in the effective 
Born--Infeld Lagrangian invariant under the shift $\,(\tx F,B)\too(\tx
F+\sfd\widetilde\Pi,B-\sfd\widetilde\Pi)$.

We remark in passing that in view of the fact that the distinguished gerbe
$\,\xcG_H\,$ is prone to arise in \emph{any} torsion geometry it would be
interesting to study the physical nature of the lower-rank components of
the local torsion gerbe data, i.e. the connection $\,(\a_{ij})_{i,j\in\cI}\,$
and the transition functions $\,(g_{ijk})_{i,j,k\in\cI}$,\, from the point
of view of the theory of gravity (to our knowledge, no such study has been
performed so far).
\bigskip

In the case of our immediate interest, i.e. for $\,\cM=G\,$ a compact simple
and simply connected Lie group, and for $\,\xcD_\a\,$ given by the conjugacy
classes
\qq\nn
\cC_\la=\left\{\ h\,\ee^{\frac{2\pi\sfi\la}{\sfk}}\,h^{-1}\quad\vert\quad
h\in G\ \right\}\,,\qquad\qquad\la\in\faff{\ggt}
\qqq
of $\,G\,$ wrapped by stable (untwisted) maximally symmetric WZW D-branes
(or their one-sided $\,G$-translates), the gauge group of each $\,\cE_\a$
could a priori be arbitrary, as long as it contains a distinguished $\,U(1)\,$
subgroup. Indeed, it was shown in \cite{gawreis,gawAb} that the gauge bundle
reduces to a (direct sum) of twisted $\,U(1)\,$ gauge bundles for any (stack
of) untwisted maximally symmetric D-brane(s) on a simply connected Lie-group
manifold $\,G$.\, For $\,\xcD_\a\equiv\cC_{\la_\a}\,$ as above, the field
strength of the corresponding twisted gauge field is given by
\qq\label{Aback}
\cF\lb h\,\ee^{\frac{2\pi\sfi\la_\a}{\sfk}}\,h^{-1}\rb=\frac{\sfk}{4\pi}\,
\tr\,\left[(h^{-1}\,\sfd h)\,\ee^{\frac{2\pi\sfi\la_\a}{\sfk}}\,(h^{-1}\,
\sfd h)\,\ee^{-\frac{2\pi\sfi\la_\a}{\sfk}}\right]\ox\bd1_N
\qqq
and is globally defined and Abelian. Granted the existence of a $\,U(1)\,$
component of the gauge group of the twisted gauge bundle, we may readily
convince ourselves that the twisted gauge field splits into a flat untwisted
(possibly) non-Abelian component and a twisted $\,U(1)\,$ component with
the curvature defining $\,\xcD_\a\,$ as above. 
\bigskip

From our discussion so far, there emerges a dual interpretation of the Kalb--Ramond
field $\,B\,$ -- on the one hand (Sect.\,\ref{sub:torgeo}), we identify it
as a torsion potential, that is a gauge field for the tangent-space translational
symmetry (of the $\si$-model), with the simple transformation property
\eqref{Bastor} under tangential coordinate shifts; on the other hand (Sect.\,\ref{sub:hola}),
it reappears as part of a rich differential-geometric structure of a bundle
gerbe $\,\cG_H\xrightarrow[loc.]{}(B,\a,g)\,$ with the globally defined (closed)
3-form field $\,H\,$ as its curvature, induced in the target manifold of
the $\si$-model due to the presence of the topological WZ term. Accordingly,
the equivalence relations \eqref{bgDeq}, expressing the notion of a stable
isomorphism of bundle gerbes in the local language, acquire the geometric
interpretation of gauge equivalences for the symmetry group of tangent-space
translations. The latter symmetry is in turn transmitted through the scalar
twist term into the $\,U(1)$-reducible twisted gauge theory supported by
a $\,\cG_H$-brane world-volume and thus augments the standard gauge symmetry.
In fact, this scalar-shift extension of the unitarily implemented gauge
symmetry can be thought of as the lowest-order term in the $\,\ell^{-1}\,$
expansion of the
full translational symmetry explored in \cite{rSr2007tap}.
\label{lcomment}

It should be remarked that there arises an elementary discrepancy between
the situation described here and the one encountered in the non-commutative
matrix geometry: in the smooth ($\si$-model) setting, the gauge field is
\emph{defined} as tangential to the $\,\cG_H$-brane world-volume, and by
construction one allows only those gauge transformations of the gerbe which
keep the world-sheet boundary fixed. In what follows, we shall ultimately
relax this constraint in order to account for shape fluctuations of $\,\cG_H$-brane
world-volumes, in a manner consistent with our geometric interpretation of
the gauge freedom underlying the structure of the torsion gerbe.

\section{The spectral geometry of the BCFT}\label{sec:NCGsection}

\noindent
We now turn to studying the low-energy effective action \eqref{SARS} within
the framework of non-commutative geometry. That the world-volumes of maximally
symmetric WZW branes can be viewed as quantisations of conjugacy classes
in the group target was already shown in \cite{ars1}, and the action \eqref{SARS}
indeed has the form of a gauge theory on such a quantised conjugacy class
-- taking, however, a very special form. In this section, we will first introduce
the main ingredients of the non-commutative spaces that come into play (more
details are spelled out in the Appendix) and then formulate an extended gauge
principle -- borrowed from the classical gerbe picture of WZW branes -- which
singles out the action \eqref{SARS} among a large class of actions that enjoy
``conventional'' gauge invariance. On the way, we will also remark that Weitzenb\"ock
geometry appears naturally on non-commutative WZW branes as well.

%There are several possible ways of addressing the issue of (D-brane) geometry
%emerging from models of stringy propagation in target spacetimes, ranging
%from the microscopic methods of (B)CFT based on the study of distinguished
%sectors of the stringy Hilbert space and the structure of boundary states
%all the way to the harmonic reconstruction of classical bulk and boundary
%geometries, supplemented with the analysis of D-brane stability conditions.
%Below, we choose instead to embed the said issue in the broad context of
%spectral NCG, which seems all the more to the point that the latter builds
%upon the notions of a Dirac operator and an abstract ``function'' algebra,
%both of which are inherent to supersymmetric formulations of string theory.

\subsection{The spectral data}

\noindent
We first show how to extract a spectral triple from boundary WZW models.
The construction is inspired by Fr\"ohlich's and Gaw\c{e}dzki's work \cite{frogaw}
on the non-commutative geometry of closed strings, but adapted to the boundary
case -- which in fact affords more natural choices of algebras and Hilbert
spaces than the bulk setting.

Following Connes \cite{connes}, the starting point of studying a non-commutative
space is its spectral triple consisting of an algebra $\,\cA\,$ (generalising
the algebra of suitable functions on a manifold), a generalised Dirac operator
$\,D$,\, and a Hilbert space $\,\ceH\,$ (generalising the space of square-integrable
spinors) on which both act. Quantum field theories come with algebras and
Hilbert spaces in any case, and in conformal QFT there is also a natural
candidate, namely the Virasoro mode $\,L_0$,\, for a generalised Laplace
operator (see in particular \cite{RogWend} for work exploiting this); in
superconformal field theories, there is an equally natural candidate for
the Dirac operator, namely the mode $\,G_0\,$ of the superconformal current
-- acting on the Ramond sector of the state space.

In the Appendix, we describe the structure of supersymmetric WZW models in
some detail (following \cite{fux}), here we merely list the ingredients of
the non-commutative spectral data associated to those SCFTs on world-sheets
with boundary. Note that passing to the supersymmetric version does not affect
the structure of WZW models significantly: the bosonic sector stays the same
except for the shifting of the level by the dual Coxeter number of the group,
and is tensored by decoupled free fermions.

We start from a boundary WZW model with a maximally symmetric boundary condition
labelled by $\,\la\in\faff{\ggt}$.\, As we are interested in the low-energy
effective action, we restrict to the subspace $\,\ceH^\la_{\rm NS}\,$ of
states (from the NS sector) whose conformal dimensions tend to zero in the
limit \eqref{decouplinglimit}, i.e. to NS-primaries and their descendants
under the finite-dimensional horizontal Lie (sub)algebra $\,[\kmg]_{(0)}
\cong\ggt$.\, We take the boundary fields associated to those states as our
algebra of ``functions'' $\,\cA_\la$,\, with multiplication induced by the
Operator Product Expansion (OPE) in the decoupling limit (where all
singularities disappear and where the product becomes associative) -- see
\cite{ars1}. This algebra is in fact a full matrix algebra of size equal
to the dimension of the \irrep of $\,\ggt\,$ labelled by $\,\la$;\, extension
to stacks of $\,n\,$ identical D-branes is trivial.

This algebra acts on the whole state space of the theory, but we restrict
to the R-sector -- as we also want the ``Dirac operator'' $\,G_0\,$ to act
-- and to the states of the lowest conformal dimension within that; we call
this subspace $\,\ceH^\la$.\, The spectral triple associated to our model
is therefore
\qq\nn
\cT^\la_{\rm WZW}:=(\cA^\la,D^\la_{\rm R},\ceH^\la)\,.
\qqq
The action of the algebra on the Hilbert space is compatible with its OPE-induced
product. The explicit formula for the Dirac operator can be derived from
expanding $\,G_0\,$ in terms of fermion and current modes (see the Appendix);
it reads
\qq\label{DiRac}
D_{\rm R}^\la:=-\sfi\,\sqrt{\sfk}\,G_0\big\vert_{\ceH^\la}=\g^a\ox L^\la_a
-\frac{\sfi}{12}\,f_{abc}\,\g^a\,\g^b\,\g^c\ox\bd1_{D_\la}\,, 
\qqq
where the $\,\g^a\,$ are rescaled fermion modes (satisfying the Clifford
algebra), and the tensor product is \wrt the above-mentioned factorisation
of the supersymmetric WZW model into a bosonic part and free fermions. $\,L^\la_a\,$
denotes the action of horizontal-subalgebra generators $\,\widehat{J}^a_0\,$
on the space $\,\ceH^\la_{\rm NS}\,$ of dimension $\,D_\la$,
\qq\nn
L^\la_a\lact f\equiv[\,Y^\la_a,f\,]:=\widehat{J}^a_0\,f
\qqq
for all $\,f\in\ceH^\la_{\rm NS}$,\, where $\,Y^\la_a\,$ denote the standard
linear combinations of the vector hyperharmonics of $\,\ggt\,$ compatible
with the choice of the Lie-algebra basis.

In summary, our choice of spectral data leads to a finite-dimensional matrix
geometry for the quantised brane world-volume. It seems to be the only natural
choice from the point of view of low-energy effective string theory: focussing
on low-energy modes dictates the truncation, open-string couplings via OPE
induce the algebra structure.
\smallskip

It is relatively straightforward to derive the spaces of differential forms
$\,\widetilde\Om^p_{D^\la_{\rm R}}\bigl(\cA^\la\bigr)\,$ and the action of
the exterior differential $\,\sfd\,$ from the spectral triples $\,\cT^\la_{\rm
WZW}$,\, following Connes' general construction; a similar computation has
been performed, in the closed-string setting, in \cite{frogreck,grand}. Details
are given in the Appendix; here, let us just recall that on a $\,p$-form
\qq\nn
\om=\sum_{i\in I}\,a^{(0)}_i\,\bigl[\,D^\la_{\rm R},a^{(1)}_i\,\bigr]\, 
\bigl[\,D^\la_{\rm R},a^{(2)}_i\,\bigr]\cdots\bigl[\,D^\la_{\rm R},a^{(p)
}_i\,\bigr]\,,\qquad\qquad a^{(n)}_i\in\cA^\la\,,
\qqq
the exterior differential acts as
\qq\nn
\sfd\,\om=\sum_{i\in I}\,\bigl[\,D^\la_{\rm R},a^{(0)}_i\,\bigr]\,\bigl[\,
D^\la_{\rm R},a^{(1)}_i\,\bigr]\,\bigl[\,D^\la_{\rm R},a^{(2)}_i\,\bigr]
\cdots\bigl[\,D^\la_{\rm R},a^{(p)}_i\,\bigr]\,.
\qqq
Due to the underlying group symmetry and to simplifications characteristic
for matrix geometries \cite{madore,gratus}, the spaces of forms can be written
rather neatly as
\qq\nn
\widetilde\Om^p_{D^\la_{\rm R}}\bigl(\cA^\la\bigr)\cong\left\{\ \g^{a_1 a_2
\ldots a_p}\ox f_{a_1 a_2\ldots a_p}\quad\big\vert\quad f_{a_1 a_2\ldots
a_p}\in\cA^\la\ \right\}\,,
\qqq
where the $\,\g^{a_1 a_2\ldots a_p}\,$ are totally antisymmetric products
of $\,\g^a$.

It is worth noting that the differential calculus thus constructed is naturally
$\,d$-dimen\-sio\-nal. In D-brane language, this means that it does not allow
for a gauge-covariant split of the gauge-field 1-forms $\,A\,$ into components
tangent to the world-volume $\,\cC_\la\,$ (this would represent a ``bona
fide'' gauge field) and those normal to it (representing transverse scalars).

Our finite-dimensional calculus allows to define a Hodge star operator $\,\star_H$,\,
mapping $\,p$-forms to $\,(d-p)$-forms: As shown in \cite{madore}, the Hodge
operator can be chosen in precisely the same form as in the commutative case;
with our flat metric $\,\d_{ab}$,\, the Hodge star $\,\star_H\,$ simply amounts to contracting
form indices with the totally antisymmetric tensor of rank $\,d$. 
\medskip

We will need differential forms in our study of gauge-invariant actions on
the quantised world-volume below. Before turning to that, let us pause to
characterise the geometry associated to the spectral data $\,\cT^\la_{\rm
WZW}\,$ further. On the free $\,\cA^\la$-module $\,\widetilde\Om^1_{D^\la_{\rm
R}}$,\, there is a natural definition of a connection \cite{frogaw}, namely
\qq\nn
\nabla_{\rm R}^\la\ :\ \widetilde\Om^1_{D^\la_{\rm R}}\bigl(\cA^\la\bigr)
\ni\om\too\nabla_{\rm R}^\la\,\om:=\g^a\ox\lb\sfi\,L_a^\la\lact\om\rb\in
\widetilde\Om^1_{D^\la_{\rm R}}\bigl(\cA^\la\bigr)\ox_{\cA^\la}\widetilde
\Om^1_{D^\la_{\rm R}}\bigl(\cA^\la\bigr)\,.
\qqq
As was first shown in the closed-string setting in \cite{frogaw}, this connection
has two fundamental properties: First, it parallelises the fuzzy geometry
at hand,
\qq\nn
\nabla_{\rm R}^\la\lb\g^a\ox\bd1_{\dim\ceH^\la_{\rm NS}}\rb=0\,,
\qqq
and hence has a vanishing curvature,
\qq\nn
\cR(\nabla_{\rm R}^\la)\equiv-\lb\nabla_{\rm R}^\la\rb^2\bigg\vert_{\widetilde
\Om^1_{D^\la_{\rm R}}\lb\cA^\la\rb}=0\,.
\qqq
Secondly, it has a non-vanishing torsion
\qq\nn
\cT(\nabla_{\rm R}^\la)\lb\g^a\ox\bd1_{\dim\ceH^\la_{\rm NS}}\rb\equiv  
\left[\lb\sfd-m\circ\nabla_{\rm R}^\la\rb\lb\g^a\ox\bd1_{\dim\ceH^\la_{\rm
NS}}\rb\right]_{\tx{2-form}}=\frac{1}{2}\,f_{abc}\,\g^{bc}\ox\bd1_{\dim 
\ceH^\la_{\rm NS}}\,,
\qqq
where $\,m\ :\ \widetilde\Om^1_{D^\la_{\rm R}}\bigl(\cA^\la\bigr)\ox_{\cA^\la}
\widetilde\Om^1_{D^\la_{\rm R}}\bigl(\cA^\la\bigr)\too\widetilde\Om^2_{D^\la_{\rm
R}}\bigl(\cA^\la\bigr)\,$ is the natural projection map (which eliminates
the ``junk forms'', see the Appendix). In other words, we find the characteristic
properties of a (non-commutative) Weitzenb\"ock geometry! We are led to conclude
that the effective fuzzy D-brane geometry inherits crucial features of its
smooth, classical counterpart, where the Cartan--Killing metric and the Kalb--Ramond
3-form conspire in such a way that the corresponding connection is
flat and torsion-full. The associated notion of teleparallelism, with its
principle of gauging tangential translations, will be invoked and exploited
in the next subsection.

\subsection{Non-commutative gauge symmetries: ordinary and extended}

\noindent
Having reconstructed the differential calculus from the spectral triple $\,\cT^\la_{\rm
WZW}$,\, we may next consider the most natural gauge dynamics based on it:
We build a field-strength 2-form
\qq\label{Fclass}
F(A)=[\,\sfd A+A\wedge A\,]_{\tx{2-form}}
\qqq
(the subscript refers to projecting out ``junk forms''), and use this to
obtain the (non-commutative) Yang--Mills action functional 
\qq\label{YMNCG}
\cS_{\rm YM}[A]\equiv\frac{1}{4\,g_{\rm YM}^2}\,\Tr_{\ceH^\la}\,\bigl[\,F(A)
\wedge\star_H F(A)\,\bigr]\,,
\qqq
where $\,\Tr_{\ceH^\la}\equiv\frac{2}{D}\,\tr_{\Om_{\rm R}}\ox
\tr_{\ceH^\la_{\rm NS}}$,\, with $\,D\,$ the dimension of the Clifford module,
and $\,g_{\rm YM}\,$ is the coupling constant of the gauge theory. The Clifford
algebra \eqref{Cliff} satisfied by $\,\g^a\,$ then enables us to rewrite
the above expression in the standard index notation
\qq\nn
\cS_{\rm YM}[A]=\frac{1}{4\,g_{\rm YM}^2}\,\tr_{\ceH^\la_{\rm NS}}\,\left[\,
F_{ab}(A)\,F_{ab}(A)\,\right]\,,
\qqq
where - by virtue of \eqref{Fclass} - the field strength takes the form familiar
from \eqref{FFzz}, see the Appendix for some details.

Since $\,F_{ab}\,$ contains a contribution $\,f_{abc}\,A_c$,\, the Yang--Mills
action \eqref{YMNCG} actually has a mass term for the gauge field. Nevertheless,
it is the simplest and most natural action invariant under the action of
the gauge group \cite{connes}
\qq\nn
\cU(\cA^\la):=\left\{\ u\in\cA^\la\quad\big\vert\quad u\,u^*=\bd1_{D_\la}
=u^*\,u\ \right\}\cong U(D_\la)
\qqq
under which the gauge field transforms as usual
\qq\nn
A\too A^u:=u^*\,A\,u+\sfi\,u^*\sfd u\,.
\qqq

Thus far, everything is standard in non-commutative geometry. One might,
however, notice that, due to the defining identity for torsion,
\qq\nn
\sfd\equiv m\circ\nabla^\la_{\rm R}+\cT(\nabla^\la_{\rm R})\,,
\qqq
the Yang--Mills action \eqref{YMNCG} can be viewed as a model of a gauge
field $\,A\,$ in a (non-commutative) Weitzenb\"ock geometry, which is coupled
``minimally'' to the torsion field\footnote{Similar couplings have been indeed
considered in classical teleparallel gravity, cf. \cite{torel}.}. This observation,
taken together with the identification of the gauge field $\,A\,$ as an
excitation of the \emph{twisted} background \eqref{Aback}, suggests to look for a non-commutative implementation
of gerbe structures, in particular of the translational symmetry. To this end, we propose to consider the non-commutative analogue
\qq\label{cons}
A^{u,\Pi}=A^u+\Pi
\qqq
of the classical transformation behaviour \eqref{APik} of the gauge field.
(In view of the decomposition $\,X_a=Y_a+A_a$,\, the non-commutative gauge
field $\,A_a\,$ scales like a covariant coordinate, while the classical gauge
field from  \eqref{APik} scales like a covariant derivative; the relation
between the two has been worked out for flat backgrounds in \cite{seiwitt-nc}
and involves the $\,B$-field in just the same way as the passage from $\,\Pi\,$
to $\,\widetilde\Pi\,$ in \eqref{compareB}. This explains why in \eqref{cons}
we see $\,\Pi\,$ rather than $\,\widetilde\Pi$.)

We restrict the 1-form $\,\Pi\,$ to be of the form
\qq\nn
\Pi=\g^a\ox\pi_a\cdot\bd1_{D_\la}\,,\qquad\qquad{\rm with}\quad\pi_a\in 
\bC\,,
\qqq
which is necessary and sufficient for the two symmetries (unitary gauge transformations
and scalar shifts) to commute with one another. This is in turn motivated
by the results of Sect.\,\ref{sub:torgeo} in which we identified world-sheet-constant
tangential translations as a remnant of left-translations in the group, independent
of the more familiar adjoint group action\footnote{But see comments at the
bottom of p.\
\pageref{lcomment}.}. Thus, our extended gauge symmetry group is $\cU(\cA^\la)\x \bC^d$. Eq.\ \eqref{cons} implies the simple transformation
behaviour
\qq\nn
F^{u,\Pi}=F^u+\sfd\Pi\,,\qquad\qquad F^u=u^*\,F\,u
\qqq
for the gauge-field strength under the extended symmetry -- compare to \eqref{gaugecltw}
and \eqref{compareB} in the bundle-gerbe context. 

Note that in the case of flat D-branes carrying a constant Kalb--Ramond field
(with $\,H=0\,$), one has $\,\sfd\Pi=0\,$ so that $\,F\,$ is invariant under
shifts; our extension of the unitary symmetry is non-trivial only for curved
backgrounds.
This is well in keeping with the observation that the Yang--Mills term of
the matrix model for flat D-branes does not receive any corrections, to leading
order of the perturbative expansion.

In the WZW case, on the other hand, postulating invariance under extended
gauge transformations allows us to derive non-trivial corrections to the
Yang--Mills action. Thus, we introduce the corrected action
\qq\nn
\cS_\cE[A]=\cS_{\rm YM}[A]+\D_\cE\cS[A]
\qqq
and demand the relation
\qq\label{extdef}
\cS_{\rm YM}[A^{u,\Pi}]+\D_\cE\cS[A^{u,\Pi}]\must\cS_{\rm YM}[A]+\D_\cE\cS[A]
\qqq
to hold for $\,A^{u,\Pi}\,$ as above. Among admissible extensions $\,\D_\cE\cS\,$
of the non-commutative Yang--Mills functional $\,\cS_{\rm YM}\,$ for untwisted
gauge fields, there is a distinguished class -- the minimal ones, with at
most two derivatives of the gauge field. In what follows, we restrict our
search to such minimal extensions.

Any extension satisfying \eqref{extdef} should in particular provide, upon
a scalar variation $\,A\too A+\Pi$,\, a counterterm for the part of the variation
of the Yang--Mills functional that is quadratic in $\,\Pi$,
\qq\label{quadlate}
\bigl(\,\d_\Pi\cS_{\rm YM}[A]\,\bigr)_{\pi^2}=\frac{\gvee N}{2\,g_{\rm YM}^2}
\,\d^{ab}\,\pi_a\,\pi_b\,.
\qqq
In the present case, the most systematic method of constructing a suitable
extension of the Yang--Mills action is to write down all possible expressions
quadratic in $\,A\,$ and then isolate those that both have the required index
structure \eqref{quadlate} and admit a completion (by higher-order terms
in $\,A\,$) that makes them invariant under standard unitary gauge transformations.
Listing all possible terms is much easier in a covariant notation employing
the exterior derivative, the wedge product and the Hodge star. In this notation,
the variation in \eqref{quadlate} is proportional to $\,\Pi\wedge\star_H
\Pi$. 

We arrive at twelve distinct candidates\footnote{One should keep in mind
that they all appear under the trace, which implies certain obvious equivalences
that we have taken into account.} for the quadratic counterterm in the extension
\qq
&A\wedge A\,,\qquad A\wedge\star_H A\,,&\non\non
&A\wedge\sfd A\,,\qquad A\wedge\star_H\sfd A\,,\qquad\star_H A\wedge\star_H
\sfd A\,,&\non\non
&A\wedge\star_H\sfd\star_H A\,,\qquad\star_H A\wedge\star_H\sfd\star_H A
\,,&\non\non
&\sfd A\wedge\sfd A\,,\qquad\sfd A\wedge\star_H\sfd A\,,&\non\non
&\sfd A\wedge\star_H\sfd\star_H A\,,\qquad\star_H\sfd A\wedge\star_H\sfd
\star_H A\,,\qquad\star_H\sfd\star_H A\wedge\star_H\sfd\star_H A\,.&\nonumber
\qqq
Note that we do not demand the forms to be of top degree $\,d\,$ since in
our matrix geometry the integral is replaced by the trace. Therefore, the
form degree is not directly relevant, as long as it does not exceed $\,d\,$
(which is why $\,\star_H A\wedge\star_H A\,$ does not appear in the list:
this form would have degree $\,2d-2>d\,$). 

We can exclude all but one of the above candidates: First, note that the
five terms containing the divergence $\,\star_H\sfd\star_H A\,$ differ in
their index structure from \eqref{quadlate} (written out, the divergence
of $\,A\,$ reads $\,L_a\,A_a\,$), and therefore they cannot cancel that variation.

Likewise, the expressions $\,A\wedge A,\,A\wedge\star_H\sfd A\,$ and    
$\,\sfd A\wedge\sfd A\,$ have the wrong index structures (e.g., $\,A\wedge
A\sim\varepsilon^{ab}\,A_a\,A_b\,$) and cannot be used to cancel \eqref{quadlate}
-- even though the last of these three terms admits the gauge invariant completion
$\,F\wedge F$. 

The expression $\,\sfd A\wedge\star_H\sfd A\,$ can be dropped as its completion
is just the original Yang--Mills term $\,F\wedge\ \star_H F$. 

For dimensional reasons, the term $\,\star_H A\wedge\star_H\sfd A\,$ exists
only for $\,d\leq 3$,\, i.e. on $\,SU(2)$,\, so it is not a universal extension
(actually, being a top form on $\,SU(2)$,\, it reduces to $\,A\wedge\sfd
A\,$ in this case). 

Of the remaining two candidates, $\,A\wedge\star_H A\,$ is the standard mass
term for the gauge field and shall consequently be dropped, too, despite
its right index structure: again, it has no gauge invariant completion.

The last remaining term $\,A\wedge\sfd A\,$ does have the right structure
and can be completed to a gauge invariant Chern--Simons current\footnote{We
use the symbol $\,\g^{abc}\equiv\frac{1}{3!}\,\sum_{\si\in\gt{S}_3}\,\sign(\si)
\,\g^{\si(a)}\,\g^{\si(b)}\,\g^{\si(c)}$.}
\qq\nn
K(A)=A\wedge\sfd A+\frac{2}{3}\,A\wedge A\wedge A\equiv\left[A\,\{D^\la_{\rm
R},A\}+\frac{2}{3}\,A^3\right]_{\tx{3-form}}\equiv\g^{abc}\ox CS_{abc}\,.
\qqq
Thus, after the dust has cleared, we are left with the unique minimal extension
which reads
\qq\nn
\D_\cE\cS[A]\equiv\cS_{CS}[A]=\frac{1}{g_{\rm YM}^2}\,\Tr_{\ceH^\la}\,\bigl[\,
T(\nabla^\la_{\rm R})\wedge\star_H K(A)\,\bigl]\,,
\qqq
with the fuzzy torsion 3-form
\qq\nn
T(\nabla^\la_{\rm R}):=\frac{\sfi}{3!}\,f_{abc}\,\g^{abc}\ox\bd1_N\,.
\qqq
This shows that our former interpretation of the Yang--Mills term as a model
of a gauge field coupled minimally to the torsion field just as well applies
to the Chern-Simons extension. 

As mentioned in the introduction already, it is straightforward to check
its standard unitary gauge invariance. And invariance under the scalar shift
$\,A\too A+\Pi\,$ to all orders in $\,\Pi$,\, not just quadratic, was also
observed before. 

Altogether, for a suitable choice of the coupling constant $\,g_{\rm YM}$,\,
the complete model that we obtain is just our original matrix model action, 
\qq\nn
\cS_\cE[A]=\frac{1}{4\,g_{\rm YM}^2}\,\Tr_{\ceH^\la}\,\bigl[\,F(A)\wedge
\star_H F(A)\,\bigl]+\frac{1}{g_{\rm YM}^2}\,\Tr_{\ceH^\la}\,\bigl[\,T( 
\nabla^\la_{\rm R})\wedge\star_H K(A)\,\bigl]\equiv\cS_{\rm ARS}[A]\,.
\qqq

\section{Conclusions and outlook}

\noindent
We have seen that the low-energy effective action \eqref{SARS} of open strings
in WZW models can be rederived from the basic building blocks of the non-commutative
geometry, associated via supersymmetric BCFT to maximally symmetric D-branes,
by invoking invariance under extended shift-gauge transformations as a symmetry
principle. Then the nonstandard term in the action -- the three-dimensional
Chern--Simons term with its specific relative coupling -- arises necessarily
to complement the usual Yang--Mills term.

The spectral triple contains a generalised Dirac operator whose properties
(flatness, torsion) in particular suggest a straightforward interpretation
of the matrix model as describing the dynamics of a gauge field coupled minimally
to the torsion field of a non-commutative Weitzenb\"ock (or teleparallel)
geometry.

Torsion is also the source of the extension of the standard principle of
gauge symmetry used in that derivation. While standard gauge transformations
constitute a unitary implementation of standard
target-space isometries preserved by the BCFT, the twisted nature of the
gauge field supported by WZW D-brane world-volumes is an immediate consequence
of the torsion-gerbe structure present on the WZW manifold. The gauge field
of the model is induced by the torsion gerbe and consequently exchanges some
of its degrees of freedom with the $\,B$-field, via its non-trivial transformation
properties under the gauge shifts of the Kalb--Ramond torsion potential.
We expect this mechanism to play a similar r\^ole for all $\si$-models
with curved targets.

We have also shown that the low-energy effective action can be obtained via
the canonical quantisation of the linearised version of the underlying WZW
$\si$-model rewritten in the Schild gauge, based on a generic Poisson
structure on the (Euclidean) world-sheet.

The derivation naturally identifies D0-branes as the elementary degrees of
freedom of the matrix model -- an interpretation which conforms with the
world-sheet RG analysis, showing that higher-dimensional branes can be
viewed as bound states of D0-branes.

That it is at all possible to derive the low-energy effective action for
open strings in group manifolds from the linearised (bulk)
$\si$-model action remains mysterious to us. One might argue that, after
all, WZW models are relatively close to free theories, but we feel that the
results of Sect.\,\ref{sec:sigma} should hold in greater generality, and
that there is some deeper underlying principle to be uncovered. In the flat
case, the transition from the Schild action to a matrix model leads to the
IKKT model and was argued \cite{yon} to provide a simple implementation of
the universal stringy uncertainty relation. At the technical level, the rederivation
of the fuzzy matrix model from the linearised WZW action is largely reminiscent\footnote{We
thank K.~Gaw\c{e}dzki for drawing our attention to this analogy.} of the
methods employed in the computation of effective actions for non-linear $\si$-models
\cite{fradtsey} in terms of Riemann normal coordinates (see also \cite{friedan,princ}
and -- in a more general context -- \cite{V,BarV,dW}), with world-sheet quantisation
playing the r\^ole of a natural regularisation procedure. The results of
Sect.\,\ref{sec:sigma} certainly call for further investigation and clarification
of the physics behind them.

A more profound understanding of the structures described in the present paper is
achieved in \cite{rSr2007tap} where in particular the extended gauge symmetry 
is studied in the so-called geometric framework of gerbe theory \cite{murray,murrays}.
This allows -- from first principles -- to identify the shift symmetry \eqref{APik}
as the first-order term (in $\,\ell$, or $\,\a'\,$) of a complete expression.
In addition, it affords
a generalisation of our results
to the case of twisted maximally symmetric D-branes.
\medskip

There are a number of more concrete open questions to study. An obvious
idea is to use the complete symmetry from \cite{rSr2007tap} to determine
which higher-order corrections (in $\,\a'\,$ and the gauge field) to the
matrix model are allowed;  the result should then be compared to higher-order terms in the effective action obtained from boundary CFT. The latter
have not been calculated as yet, but the computations of \cite{cornalba,Bordalo:2004xg}
should be useful here. Another interesting issue is that of lifting the gerbe
symmetry discussed in our paper to the strictly non-classical r\'egime, and
in particular to the $\,\qgt$-related\footnote{Here $\,\qgt\,$ denotes the
Drinfel'd--Jimbo algebra with the deformation parameter $\,q=\ee^{\frac{\pi
\sfi}{\sfk+\gvee}}$,\, as suggested by the quantum symmetries
of the (B)CFT structures.} geometries of \cite{jpstein,fusion} -- one would
hope that our findings shed some light on the still unresolved problem of
defining a $\,\qgt$-covariant gauge field theory capturing the dynamics of
quantum WZW branes beyond the semiclassical approximation of the fuzzy matrix
model. It should also be worthwhile to combine our findings with the investigations
in \cite{ydri,scale} and to see what r\^ole torsion geometry plays in continuum
limits of the matrix model. 

Extension beyond models of the WZW type considered may well be tractable:
First of all, the equivalence between the Polyakov action and the Schild action can be
established for arbitrary (dilaton-free) non-linear $\si$-models \cite{rSr2007tap}
and hence allows to associate matrix models to any such background. Cosets
are a particularly important class, from the point of view of string theory.
In this context, it should be noted that the Dirac operator \eqref{DiRac}
is a special case of a more general construct known as Kostant's cubic Dirac
operator \cite{kost}, see also \cite{brink}, which was studied extensively
in the mathematical literature. The general form of this operator given in \cite{kost} coincides with the Ramond--Dirac operator derived
in \cite{frogaw} for supersymmetric cosets of WZW models. Thus, there is
a natural starting point of a discussion of effective D-brane geometry and
gauge dynamics in the NCG framework for coset $\si$-models. We hope to
return to this question in the near future.

Scalar shifts of gauge fields such as those appearing in our extended gauge
symmetries also occur in the context of Morita self-equivalences of the underlying
NCG, see \cite{CDS,seiwitt-nc,sch,bmz,ksch,psch}. Thus one might speculate
that there is some hidden T-duality group acting on the gerbe data. A related
issue is that of the (stable-isomorphism) equivalences among gerbe data:
bearing in mind that it is the Kalb--Ramond potential $\,B\,$ that defines
the non-commutative deformation of the algebra of functions on the D-brane
world-volume, it is tempting to relate the ambiguities in its definition
(which show up whenever there is a non-vanishing torsion) to some kind of
Seiberg--Witten equivalences \cite{seiwitt-nc} among various formulations
of the gauge field theory on the D-brane, differing in the choice of the
non-commutativity parameter. In contrast to the flat case, the kinetic and
the topological term in the WZW model action are linked by requiring conformal
invariance, so we would at most expect a very restricted set of equivalences.

\acknowledgments

\noindent
We are indebted to Krzysztof Gaw\c{e}dzki for numerous helpful discussions
and for generously sharing his insights with us. We are also grateful to
Jacek Pawe\l czyk for discussions and his sustained interest in our work.

This work was supported in part by the EU Research Training Network grants
`Euclid', contract number HPRN-CT-2002-00325, `Superstring Theory', contract
number MRTN-CT-2004-512194, by the PPARC rolling grant PP/C507145/1, and
by the EU Marie Curie Training Site `Strings, Branes and Boundary Conformal
Field Theory', contract number HPMT-CT-2001-00296.

\appendix
\section{Appendix}

\noindent
In the following, we collect some details on supersymmetric WZW models (a
very nice exposition is given in \cite{fux} which we largely follow), on
spectral data in non-commutative geometry (see \cite{connes} and also \cite{frogreck}),
and on how to extract the latter from the former.

Our starting point is the supercurrent of the supersymmetric WZW
model ($\,Z:=(z,\th)\,$ is the supercoordinate)
\qq\label{GinsT}
\xcT(Z):=\frac{1}{2}\,G(z)+\th\,T(z)\,,
\qqq
containing the superconformal current $\,G\,$ and the energy momentum tensor
$\,T\,$ as its superpartner. When written in terms of the super-Kac--Moody
current
\qq\label{Ja}
J^a(Z):=\sqrt{2\sfk}\,\psi^a(z)+\th\,j^a(z)
\qqq
and its superderivative ($\,\xcD:=\tfrac{\p}{\p\th}+\th\,\p,\ \p=\tfrac{\p}{\p
z}\,$)
\qq\label{DJa}
\xcD J^a(Z)=j^a(z)+\th\,\sqrt{2\sfk}\,\p\psi^a(z)\,, 
\qqq
the definition reads (the colons denote normal ordering)
\qq\label{sT}
\xcT(Z):=-\frac{1}{\sfk}\,\d_{ab}\,:\xcD J^a\,J^b:(Z)+\frac{2}{3\sfk^2}\,
f_{abc}\,:J^a\,:J^b\,J^c:\,:(Z)+\xcA\,,
\qqq
where $\,\xcA=0$ in the NS sector, but in  the R sector one has to add
the Ramond vacuum energy $\xcA=\frac{c}{16}\,Z^{-\frac{3}{2}}$,\, with $\,
Z^{-\frac{3}{2}}=\th\,z^{-2}$. 

Above, the $\,\psi^a\,$ are the fermions of the model, furnishing the adjoint
\rep of the current-symmetry algebra generated by the currents $\,j^a,\,
a=1,2,\ldots,d$.\, They satisfy the OPE
\qq
j^a(z_1)\,j^b(z_2)&\sim&-\frac{\sfk\,\d_{ab}}{2(z_1-z_2)^2}+\frac{f_{ab 
c}}{z_1-z_2}\,j^c(z_2)\,,\non\non
\psi^a(z_1)\,\psi^b(z_2)&\sim&\frac{\d_{ab}}{z_1-z_2}\,,\label{opsi}\\\non
j^a(z_1)\,\psi^b(z_2)&\sim&\frac{f_{abc}}{z_1-z_2}\,\psi^c(z_2)\,,\nonumber
\qqq
which simplifies considerably upon redefining the currents as
\qq\label{curred}
j^a\too\widehat{j}^a:=j^a+\frac{1}{2}\,f_{abc}\,:\psi^b\,\psi^c:\,.
\qqq
Indeed, for the new currents we obtain
\qq\nn
\widehat{j}^a(z_1)\,\widehat{j}^b(z_2)&\sim&-\frac{\widehat{\sfk}\,\d_{a
b}}{2(z_1-z_2)^2}+\frac{f_{abc}}{z_1-z_2}\,\widehat{j}^c(z_2)\,,\non\non
\widehat{j}^a(z_1)\,\psi^b(z_2)&\sim&0\,,\nonumber
\qqq
with the shifted value of the level $\,\widehat{\sfk}:=\sfk-g^\vee$,\, and
the original quantum field theory dissolves into a pair of mutually independent
subtheories: a bosonic $\,\widehat{\ggt}_{\widehat{\sfk}}\,$ WZW model and
the theory of $\,d\,$ free Fermi fields $\,\psi^a$.

Upon substituting \eqref{Ja} and \eqref{DJa} in \eqref{sT} and using \eqref{GinsT}
together with \eqref{curred}, we arrive at the formula
\qq\nn
-\sqrt{2\sfk}\,G(z)=\d_{ab}\,:\widehat{j}^a\,\psi^b:(z)-\frac{1}{3}\,f_{a
bc}\,:\psi^a\,:\psi^b\,\psi^c:\,:(z)\,.
\qqq
In terms of the Laurent modes
\qq\nn
G(z)=\sum_{m\in\bZ}\,z^{-m-\tfrac{3}{2}}\,G_m\,,\qquad\widehat{j}^a(z)= 
\sum_{m\in\bZ}\,z^{-m-1}\,\widehat{j}^a_m\,,\qquad\psi^a(z)=\sum_{m\in\bZ}
\,z^{-m-\tfrac{1}{2}}\,\psi^a_m\,,
\qqq
this yields an expression for the zero mode of $\,G(z)$,
\qq\label{G0}
-\sqrt{2\sfk}\,G_0=\sum_{m\in\bZ}\,\d_{ab}\,:\widehat{j}^a_m\,\psi^b_{-m}
:-\frac{1}{3}\,\sum_{m,n\in\bZ}\,f_{abc}\,:\psi^a_m\,:\psi^b_n\,\psi^c_{-
m-n}:\,:\,,
\qqq
which acts in the Ramond sector and is crucial for the NCG interpretation.
\smallskip

As described in Sect.\,\ref{sec:NCGsection} above, the Hilbert space    
$\,\ceH^\la\,$ of the spectral data associated to a maximally symmetric boundary
state $\,\xcD_\la\,$ of weight label $\,\la\in\faff{\ggt}\,$ is chosen to
be the subspace of the boundary CFT consisting of the Ramond vacuum sector
$\,\Om_{\rm R}\,$ tensored with the space $\,\ceH^\la_{\rm NS}\,$ of Neveau--Schwarz
primary boundary fields and their horizontal ($\,\left[\widehat{\ggt}_{\widehat
\sfk}\right]_{(0)}\cong\ggt\,$) descendants.

We take our algebra of ``functions'' $\,\cA^\la\,$ to be isomorphic to  
$\,\ceH^\la_{\rm NS}\,$ as a vector space (and we will make use of this freely
in the following). An element $\,a\in\cA^\la\,$ is represented on       
$\,\ceH^\la\,$ by the matrix $\,\bd1_D\ox a$,\, acting on $\,\om\ox b\in
\ceH^\la\,$ as $\,a\lact(\om\ox b):=\om\ox a\,b$,\, with the matrix product
$\,a\,b\,$ defined by the decoupling limit of the boundary OPE \cite{ars1,ars2},
which then also defines the product the algebra is endowed with.

We want to restrict \eqref{G0} to the subspace $\,\ceH^\la\,$ of minimal-energy
Ramond states and use it as the generalised Dirac operator of our spectral
triple. We first recall that the Ramond vacuum fields $\,\si_i,\,i=1,2, 
\ldots,D:=2^{\left[\frac{d+1}{2}\right]}\,$ can be generated from the spin
field $\,\si\equiv\si_1\,$ of \cite{fux} by the action of the free-fermion
zero-mode OPE subalgebra $\,\psi^a_0$,\, and that they satisfy
\qq\nn
%\forall_{m \in \bN \setminus \{0\}} \ \forall_{i = 1,\ldots,D} \ : \
j^a_m\,\si_i=0=\psi^a_m\,\si_i
\qqq
for $\,m\neq 0$.\, Using this, it is not difficult to see that the restriction
of the operator \eqref{G0} to $\,\ceH^\la:=\Om_R\ox\ceH^\la_{\rm NS}\,$ yields
\qq\nn
-\sfi\,\sqrt{\frac{\sfk}{2}}\,G_0\big\vert_{\ceH^\la}=\frac{\sfi}{2}\,  
\d_{ab}\,\psi^a_0\ox\widehat{j}^b_0-\frac{\sfi}{6}\,f_{abc}\,\psi^a_0\, 
\psi^b_0\,\psi^c_0\ox\bd1_{D_\la}
\qqq
with $\,D_\la=\dim\ceH^\la_{\rm NS}$.\, We now pass to rescaled current modes
$-2\sfi\,\widehat{J}^a_0\equiv\widehat{j}^a_0\,$ and rescaled fermion zero
modes $\,\g^a:=\sqrt{2}\,\psi^a_0$,\, which due to the OPE \eqref{opsi} satisfy
the relations of the Clifford algebra $\,\Cliff\,(\bR^d)$,
\qq\label{Cliff}
\{\g^a,\g^b\}=2\,\d^{ab}\,.
\qqq
Furthermore, we introduce the notation
\qq\nn
\widehat{J}^a_0\,f=:L^\la_a\lact f
\qqq
for the action of the horizontal-subalgebra generators $\,\widehat{J}^a_0\,$
on $\,f\in\ceH^\la_{\rm NS}$,\, as dictated by the relevant boundary OPE
\qq\nn
\widehat{J}^a(x_1)\,f(x_2)\sim\frac{1}{x_1-x_2}\,L^\la_a\lact f(x_2)\,.
\qqq
By a slight abuse of notation, we will use the same symbol $\,\lact\,$ for
the action of the $\,L^\la_a\,$ on algebra elements. The $\,L^\la_a\,$ are then
self-adjoint generators of the Lie algebra $\,\ggt\,$ and accordingly satisfy
\qq\label{Lalg}
[\,L^\la_a,L^\la_b\,]=\sfi\,f_{abc}\,L^\la_c\,.
\qqq
The action of $\,L^\la_a\,$ is tantamount to an inner derivation\footnote{The
statement can readily be verified in the hyperharmonic basis of the algebra
of functions on the fuzzy conjugacy class $\,\xcD_\la$.\, First, it is elementary
to check that \eqref{Lasx} yields an action of $\,\ggt\,$ on the vector multiplet
itself (the coordinate ``functions''), concordant with the representation
label carried by the latter. Then, all we need to extend this action to homogeneous
polynomials in the coordinates, reproducing higher hyperharmonics, is the
primitive coalgebra structure on $\,\ggt$.}
\qq\label{Lasx}
L^\la_a\lact f=[\,Y^\la_a,f\,]\,,
\qqq
where $\,Y^\la_a\,$ denote the standard linear combinations of the vector
hyperharmonics of $\,\ggt\,$ compatible with the choice of the basis of $\,\ggt\,$
in \eqref{Lalg}.

This leads to the following expression for the generalised Dirac operator
of the spectral triple (cp \cite{frogaw})
\qq\label{ADiRac}
D_{\rm R}^\la:=-\sfi\,\sqrt{\sfk}\,G_0\big\vert_{\ceH^\la}=\g^a\ox L^\la_a
-\frac{\sfi}{12}\,f_{abc}\,\g^a\,\g^b\,\g^c\ox\bd1_{D_\la}\,.
\qqq
Note the presence of a torsion term $\,\frac{\sfi}{12}\,f_{abc}\,\g^a\, 
\g^b\,\g^c\ox\bd1_{D_\la}$.
\medskip

We have now established the data of the spectral triple $\,\cT^\la_{\rm WZW}
=(D^\la_{\rm R},\cA^\la,\ceH^\la)\,$ that we associate to the boundary WZW model,
and will in the following work out those elements of the differential calculus
of $\,\cT^\la_{\rm WZW}\,$ which are instrumental for the main body of this
paper. The intrinsic definition of a spectral differential calculus is due
to Connes \cite{connes}, many details are e.g. given in \cite{frogreck}.
We are mainly interested in differential $\,p$-forms on the non-commutative
D-brane world-volumes, in particular for $\,p=1\,$ (the gauge field 1-form)
and $\,p=2\,$ (the gauge field strength 2-form). On a (representative) of
a $\,p$-form
\qq\nn
\om=\sum_{i\in I}\,a^{(0)}_i\,[\,D^\la_{\rm R},a^{(1)}_i\,]\,[\,D^\la_{\rm
R},a^{(2)}_i\,]\cdots[\,D^\la_{\rm R},a^{(p)}_i\,]
\qqq
($\,I\,$ is some index set, $\,a^{(n)}_i\in\cA^\la\,$), the standard definition
of the exterior derivative is
\qq\nn
\sfd\ :\ \om\too\sfd\,\om\equiv D^\la_{\rm R}\,\om-(-1)^p\,\om\,D^\la_{\rm
R}
\qqq
and reduces to the explicit formula
\qq\nn
\sfd\,\om=\sum_{i\in I}\,[\,D^\la_{\rm R},a^{(0)}_i\,]\,[\,D^\la_{\rm R},
a^{(1)}_i\,]\,[\,D^\la_{\rm R},a^{(2)}_i\,]\cdots[\,D^\la_{\rm R},a^{(p)}_i
\,]\,;
\qqq
this uses the Lichnerowicz formula which relates $\,(D^\la_{\rm R})^2\,$
to the Virasoro Hamiltonian $\,L_0$,\, as explicited (in the bulk case) in
\cite{frogaw}. The definitions of differential 0-forms
\qq\nn
\widetilde\Om^0_{D^\la_{\rm R}}\lb\cA^\la\rb=\cA^\la
\qqq
and of differential 1-forms
\qq\nn
\widetilde\Om^1_{D^\la_{\rm R}}\lb\cA^\la\rb=\left\{\ \g^a\ox a_a\quad  
\bigg\vert\quad a_a=\sum_{i\in I}\,a^{(0)}_i\,\left[\,Y^\la_a,a^{(1)}_i\,
\right],\quad a^{(0),(1)}_i\in\cA^\la\ \right\}
\qqq
are straightforward, but that of differential 2-forms is somewhat more involved
\cite{connes}: From the set of all
\qq\nn
\om=\sum_{i\in I}\,a^{(0)}_i\,\left[\,\cD^\la_{\rm R},a^{(1)}_i\,\right]
\,\left[\,\cD^\la_{\rm R},a^{(2)}_i\,\right]\,,
\qqq
we have to project out the so-called ``junk forms'', i.e. operators of the
kind
\qq\label{junk2}
\eta:=\sum_{i\in I}\,\left[\,\cD^\la_{\rm R},a^{(0)}_i\,\right]\,\left[\,
\cD^\la_{\rm R},a^{(1)}_i\,\right]\qquad\tx{such that}\quad\sum_{i\in I}\,
a^{(0)}_i\,\left[\,\cD^\la_{\rm R},a^{(1)}_i\,\right]\equiv 0\,.
\qqq
Using the explicit form of the Ramond--Dirac operator \eqref{ADiRac} and
the Clifford algebra \eqref{Cliff}, we see that the junk 2-forms are 
\qq\label{eta2}
\eta=-\sum_{i\in I}\,\d^{ab}\cdot\bd1_D\ox a^{(0)}_i\,(L^\la_a\,L^\la_b 
\lact a^{(1)}_i)\,.
\qqq
%(the action $\lact$
%on $a \in \cA^\la$ is the same as the action $\lact$ on $a \in \ceH^\la_{NS}$
%and the tilde merely distinguishes between the two domains of the action).
The operator acting on $\,a^{(1)}_i\,$ in \eqref{eta2} is just the Laplacian
on the fuzzy conjugacy class associated with $\,\cD_\la$.\, This observation,
together with the fact that the ``junk forms'' furnish a left $\,\cA^\la$-module,
allows to show that
\qq\nn
\widetilde\Om^2_{D^\la_{\rm R}}\lb\cA^\la\rb=\left\{\ \g^{ab}\ox f_{ab}
\quad\bigg\vert\quad f_{ab}=\sfi\,\sum_{i\in I}\,a^{(0)}_i\,[\,Y^\la_{[a}
,a^{(1)}_i\,]\,[\,Y^\la_{b]},a^{(2)}_i\,]\,,\quad a^{(0),(1),(2)}_1\in  
\cA^\la\ \right\}\,,
\qqq
where $\,\g^{ab}:=-\frac{\sfi}{2}\,\left[\,\g^a,\g^b\,\right]$.

As a sample computation to indicate the relevance of projecting out ``junk
forms'', let us look at the $\,\sfd A$-piece of the Yang--Mills field strength
$\,F$:\, Starting from the 1-form
\qq\nn
A=\g^a\ox\sum_{i\in I}\,a^{(0)}_i\,\lb L^\la_a\lact a^{(1)}_i\rb\,,
\qqq
we compute the 2-form part of $\,\sfd A\,$ by projecting out components of
the type \eqref{junk2}, which results in the unfamiliar term $\,f_{abc}\,
A_c\,$ in $\,F_{ab}\,$ from \eqref{SARS}, using that the $\,L^\la_a\,$ furnish
a \rep of the Lie algebra $\,\ggt$,
\qq
[\sfd A]_{\tx{2-form}}&\equiv&\lb\sum_{i\in I}\,[\,\cD^\la_{\rm R},     
a^{(0)}_i\,]\,[\,\cD^\la_{\rm R},a^{(1)}_i\,]\rb_{\tx{2-form}}=\sum_{i\in
I}\,\g^{ab}\ox\sfi\,\lb L^\la_a\lact a^{(0)}_i\rb\,\lb L^\la_b\lact     
a^{(1)}_i\rb=\non\non
&=&\g^{ab}\ox\frac{1}{2}\,\lb\sfi\,L^\la_a\lact A_b-\sfi\,L^\la_b\lact A_a
+f_{abc}\,A_c\rb\,.\nonumber
\qqq
The expressions for 1-forms and 2-forms given above are still slightly unwieldy.
Arguments given in \cite{madore,gratus} for rather general non-commutative
spectral data over full matrix algebras show that the spaces of $\,p$-forms
are in fact free modules over the algebra of functions $\,\cA^\la$, 
\qq\nn
\widetilde\Om^p_{D^\la_{\rm R}}\bigl(\cA^\la\bigr)\cong\left\{\ \g^{a_1 a_2
\ldots a_p}\ox f_{a_1 a_2\ldots a_p}\quad\big\vert\quad f_{a_1 a_2\ldots
a_p}\in\cA^\la\ \right\}\,,
\qqq
where the $\,\g^{a_1 a_2\ldots a_p}\,$ are totally antisymmetric products
of $\,\g^a$.


\begin{thebibliography}{M}


\bibitem{ars1}
{\sc A.Yu.~Alekseev, A.~Recknagel, and V.~Schomerus},
``Non-commutative world-volume geometries: branes on $\,SU(2)\,$ and fuzzy
spheres", 
{\em JHEP\/} {\bf 9909} (1999) 023
[\href{http://arXiv.org/abs/hep-th/9908040}{{\tt hep-th/9908040}}].


\bibitem{ars2}
{\sc A.Yu.~Alekseev, A.~Recknagel, and V.~Schomerus},
``Brane dynamics in background fluxes and non-commutative geometry",
{\em JHEP\/} {\bf 0005} (2000) 010
[\href{http://arXiv.org/abs/hep-th/0003187}{{\tt hep-th/0003187}}].


\bibitem{ikkt}
{\sc N.~Ishibashi, H.~Kawai, Y.~Kitazawa, and A.~Tsuchiya},
``A large-$N\,$ reduced model as superstring",
{\em Nucl.\ Phys.\/} {\bf B498} (1997) 467--491
[\href{http://arXiv.org/abs/hep-th/9612115}{{\tt hep-th/9612115}}].


\bibitem{yon}
{\sc T.~Yoneya},
``Schild action and space-time uncertainty principle in string theory",
{\em Prog.\ Theor.\ Phys.\/} {\bf 97} (1997) 949--962
[\href{http://arXiv.org/abs/hep-th/9703078}{{\tt hep-th/9703078}}];
``D-particles, D-instantons, and a space-time uncertainty principle in string
theory", 
in {\em Seoul 1997, Recent developments in nonperturbative quantum field
theory}, Y.M.~Chop and M.~Virasoro (eds.), World Scientific, 1998
[\href{http://arXiv.org/abs/hep-th/9707002}{{\tt hep-th/9707002}}].


\bibitem{AleksSchom}
{\sc A.Yu.~ Alekseev and V.~Schomerus}, 
``D-branes in the WZW model",
{\em Phys.\ Rev.\/} {\bf D60} (1999) 061901
[\href{http://arXiv.org/abs/hep-th/9812193}{{\tt hep-th/9812193}}].


\bibitem{ga}
{\sc K.~Gaw\c{e}dzki}, 
"Conformal field theory: a case study"
[\href{http://arXiv.org/abs/hep-th/9904145}{{\tt hep-th/9904145}}].


\bibitem{klim}
{\sc C.~Klim\v c\'ik},
``A nonperturbative regularization of the supersymmetric Schwinger model",
{\em Commun.\ Math.\ Phys.\/} {\bf 206} (1999) 567--586
[\href{http://arXiv.org/abs/hep-th/9903112}{{\tt hep-th/9903112}}].



\bibitem{myers}
{\sc R.~C.~Myers},
``Dielectric-branes",
{\em JHEP\/} {\bf 9912} (1999) 022
[\href{http://arXiv.org/abs/hep-th/9910053}{{\tt hep-th/9910053}}].


\bibitem{flux-stab}
{\sc C.~Bachas, M.R.~Douglas, and C.~Schweigert},
``Flux stabilization of D-branes",
{\em JHEP\/} {\bf 0005} (2000) 048
[\href{http://arXiv.org/abs/hep-th/0003037}{{\tt hep-th/0003037}}].


\bibitem{paw-stab}
{\sc J.~Pawe\l czyk},
``$\,SU(2)\,$ WZW D-branes and their non-commutative geometry from DBI action",
{\em JHEP\/} {\bf 0008} (2000) 006
[\href{http://arXiv.org/abs/hep-th/0003057}{{\tt hep-th/0003057}}].


\bibitem{brs-stab}
{\sc P.~Bordalo, S.~Ribault, and C.~Schweigert},
``Flux stabilization in compact groups",
{\em JHEP\/} {\bf 0110} (2001) 036
[\href{http://arXiv.org/abs/hep-th/0108201}{{\tt hep-th/0108201}}].


\bibitem{Fred}
{\sc S.~Fredenhagen},
``Dynamics of D-branes in curved backgrounds",
Ph.D. thesis, Humboldt Universit\"at, 2002
[\href{http://edoc.hu-berlin.de/dissertationen/fredenhagen-stefan-2002-09-16/PDF/Fredenhagen.pdf}
{\tt http://edoc.hu-berlin.de/dissertationen/
\newline
\phantom{xxxxxxxxxxxxxxxxxxxxxxxx}fredenhagen-stefan-2002-09-16/PDF/Fredenhagen.pdf}].


\bibitem{StefanVolker}
{\sc S.~Fredenhagen and V.~Schomerus},
``Branes on group manifolds, gluon condensates, and twisted K-theory",
{\em JHEP} {\bf 0104} (2001) 007
[\href{http://arXiv.org/abs/hep-th/0012164}{{\tt hep-th/0012164}}].

\bibitem{Antonetal}
{\sc A.Yu.~Alekseev, S.~Fredenhagen, T.~Quella, and V.~Schomerus},
``Non-commutative gauge theory of twisted D-branes'',
{\em Nucl.\ Phys.\ } {\bf B646} (2002) 127
[\href{http://arXiv.org/abs/hep-th/0205123}{{\tt hep-th/0205123}}].

\bibitem{Monnier}
{\sc S.~Monnier},
``D-branes in Lie groups of rank $\,>1$",
{\em JHEP\/} {\bf 0508} (2005) 062
[\href{http://arXiv.org/abs/hep-th/0507159}{{\tt hep-th/0507159}}].


\bibitem{AfflLud}
{\sc I.~Affleck and A.W.W.~Ludwig},
``The Kondo effect, conformal field theory and fusion rules",
{\em Nucl.\ Phys.\/} {\bf B352} (1991) 849--862;
``Critical theory of overscreened Kondo fixed points",
{\em Nucl.\ Phys.\/} {\bf B60} (1991) 641--696;
``Universal noninteger 'ground state degeneracy' in critical quantum
systems", 
{\em Phys.\ Rev.\ Lett.\/} {\bf 67} (1991) 161--164.


\bibitem{schild}
{\sc A.~Schild},
``Classical null strings",
{\em Phys.\ Rev.\/} {\bf D16} (1977) 1722--1726.

\bibitem{rSr2007tap}
{\sc R.R.~Suszek},
``From higher to twisted gauge fields via deformed $\si$-models",
in preparation.


\bibitem{hoppe}
{\sc J.R.~Hoppe},
``Quantum theory of a massless relativistic surface and a two-dimensional
bound state problem",
Ph.D. thesis, Massachusetts Institute of Technology, 1982.


\bibitem{solo}
{\sc O.A.~Soloviev},
``On the Schild action for $\,D=0\,$ and $\,D=1\,$ strings",
{\em Mod.\ Phys.\ Lett.\/} {\bf A13} (1998) 2415--2426
[\href{http://arXiv.org/abs/hep-th/9708021}{{\tt hep-th/9708021}}].


\bibitem{bracurzach}
{\sc E.~Braaten, T.L.~Curtright, and C.K.~Zachos},
``Torsion and geometrostasis in nonlinear sigma models",
{\em Nucl.\ Phys.\/} {\bf B260} (1985) 630--688.


\bibitem{cornalba}
{\sc L.~Cornalba and R.~Schiappa},
``Nonassociative star product deformations for D-brane world-volumes in
curved backgrounds",
{\em Commun.\ Math.\ Phys.\/}  {\bf 225} (2002) 33--66
[\href{http://arxiv.org/abs/hep-th/0101219}{{\tt hep-th/0101219}}].


\bibitem{Bordalo:2004xg}
{\sc P.~Bordalo, L.~Cornalba, and R.~Schiappa},
``Towards quantum dielectric branes: Curvature corrections in abelian  beta
function and nonabelian Born-Infeld action",
{\em Nucl.\ Phys.\/} {\bf B710} (2005) 189--254
[\href{http://arxiv.org/abs/hep-th/0409017}{{\tt hep-th/0409017}}].
  

\bibitem{curzach}
{\sc T.L.~Curtright and C.K.~Zachos},
``Geometry, topology and supersymmetry in nonlinear models",
{\em Phys.\ Rev.\ Lett.\/} {\bf 53} (1984) 1799--1801.


\bibitem{agip}
{\sc V.C.~de Andrade, L.C.T.~Guillen, and J.G.~Pereira}
``Teleparallel gravity: an overview",
in {\em The Ninth Marcel Grossmann Meeting: On Recent Developments in Theoretical
and Experimental General Relativity, Gravitation and Relativistic Field Theories,
Proceedings of the MGIX MM Meeting, The University of Rome "La Sapienza"
2--8 July 2000\/}, V.G.~Gurzadyan, R.T.~Jantzen, and R.~Ruffini (eds.), World
Scientific, 2002
[\href{http://arXiv.org/abs/gr-qc/0011087}{{\tt gr-qc/0011087}}].


\bibitem{telintro}
{\sc R.~Aldrovandi and J.G.~Pereira},
``An introduction to teleparallel gravity'', Lecture notes Universidad de Concepci\'on, 2005.


\bibitem{hehl}
{\sc F.~Gronwald and F.W.~Hehl},
``On the gauge aspects of gravity",
in {\em Quantum Gravity: International School of Cosmology and Gravitation XIV Course : 80th Birthday Dedication to Peter G. Bergmann : Erice, Italy 11-19 May, 1995\/}, P.G.~Bergmann, V.~de Sabbata, and H.-J.~Treder (eds.), World Scientific, 1996
[\href{http://arXiv.org/abs/gr-qc/9602013}{{\tt gr-qc/9602013}}].


\bibitem{gawtop}
{\sc K.~Gaw\c{e}dzki},
``Topological actions in two-dimensional quantum field theories",
in {\em Non-Perturbative Quantum Field Theory\/}, G.~'t Hooft, A.~Jaffe,
G.~Mack, P.K.~Mitter, and R.~Stora (eds.), Plenum Press, 1988.


\bibitem{gawreis}
{\sc K.~Gaw\c{e}dzki and N.~Reis},
``WZW branes and gerbes",
{\em Rev.\ Math.\ Phys.\/} {\bf 14} (2002) 1281--1334
[\href{http://arXiv.org/abs/hep-th/0205233}{{\tt hep-th/0205233}}].


\bibitem{gawAb}
{\sc K.~Gaw\c{e}dzki},
``Abelian and non-Abelian branes in WZW models and gerbes",
{\em Commun.\ Math.\ Phys.\/} {\bf 258} (2005) 23--73
[\href{http://arXiv.org/abs/hep-th/0406072}{{\tt hep-th/0406072}}].


\bibitem{alva}
{\sc O.~Alvarez},
``Topological quantization and cohomology",
{\em Commun.\ Math.\ Phys.} {\bf 100} (1985) 279--309.


\bibitem{jlB}
{\sc J.-L.~Brylinski},
{\em Loop spaces, characteristic classes and geometric quantization},
Progress in Mathematics, vol. 107,
Birkh\"auser, 1993.


\bibitem{giraud}
{\sc J.~Giraud},
{\em Cohomologie non ab\'elienne},
Grundlehren der Mathematischen Wissenschaften, vol. 179, Springer, 1971.


\bibitem{murray}
{\sc M.K.~Murray},
``Bundle gerbes",
{\em J.\ London\ Math.\ Soc.\/} {\bf 54} (1996) 403--416
[\href{http://arXiv.org/abs/dg-ga/9407015}{{\tt dg-ga/9407015}}].


\bibitem{murrays}
{\sc M.K.~Murray and D.~Stevenson},
``Bundle gerbes: stable isomorphism and local theory",
{\em J.\ London\ Math.\ Soc.\/} {\bf 62} (2000) 925--937
[\href{http://arXiv.org/abs/math.DG/9908135}{{\tt math.DG/9908135}}].


\bibitem{murson}
{\sc S.~Johnson},
``Constructions with bundle gerbes",
Ph.D. thesis, University of Adelaide, 2002
[\href{http://arXiv.org/abs/math.DG/0312175}{{\tt math.DG/0312175}}].


\bibitem{wzwitt}
{\sc E.~Witten},
``Non-abelian bosonization in two dimensions",
{\em Commun.\ Math.\ Phys.\/}  {\bf 92} (1984) 455--472.


\bibitem{kapu}
{\sc A.~Kapustin},
``D-branes in a topologically nontrivial $\,B$-field",
{\em Adv.\ Theor.\ Math.\ Phys.\/} {\bf 4} (2000) 127--154
[\href{http://arXiv.org/abs/hep-th/9909089}{{\tt hep-th/9909089}}].


\bibitem{frogaw}
{\sc J.~Fr\"ohlich and K.~Gaw\c{e}dzki},
"Conformal field theory and geometry of strings",
in {\em Vancouver 1993, Proceedings, Mathematical quantum theory I: Field
Theory and Many-Body Theory\/}, J.~Feldman, R.~Froese, and L.M.~Rosen (eds.),
CRM Proceedings \& Lecture Notes, vol. 7, American Mathematical Society,
1994
[\href{http://arXiv.org/abs/hep-th/9310187}{{\tt hep-th/9310187}}].


\bibitem{connes}
{\sc A.~Connes},
{\em Noncommutative Geometry},
Academic Press, 1994.


\bibitem{RogWend}
{\sc D.~Roggenkamp and K.~Wendland},
``Limits and degenerations of unitary conformal field theories",
{\em Commun.\ Math.\ Phys.\/}  {\bf 251} (2004) 589--643
[\href{http://arXiv.org/abs/hep-th/0308143}{{\tt hep-th/0308143}}].


\bibitem{fux}
{\sc J.~Fuchs},
``Superconformal Ward identities and the WZW model",
{\em Nucl.\ Phys.\/} {\bf B286} (1987) 455--484;
``More on the super WZW theory", {\em Nucl.\ Phys.\/} {\bf B318} (1989) 631--654.


\bibitem{frogreck}
{\sc J.~Fr\"ohlich, O.~Grandjean, and A.~Recknagel},
``Supersymmetric quantum theory and (non-commutative) differential geometry",
{\em Commun.\ Math.\ Phys.\/} {\bf 193} (1998) 527--594
[\href{http://arXiv.org/abs/hep-th/9612205}{{\tt hep-th/9612205}}];
``Supersymmetric quantum theory, non-commutative geometry, and gravitation",
in {\em NATO Advanced Study Institute: Les Houches Summer School On Theoretical
Physics, Session 64: Quantum Symmetries\/}, A.~Connes, K.~Gaw\c{e}dzki, and
J.~Zinn-Justin (eds.), North-Holland, 1998
[\href{http://arXiv.org/abs/hep-th/hep-th/9706132}{{\tt hep-th/9706132}}];
``Supersymmetric quantum theory and non-commutative geometry",
{\em Commun.\ Math.\ Phys.\/} {\bf 203} (1999) 119--184
[\href{http://arXiv.org/abs/hep-th/hep-th/hep-th/9807006}{{\tt hep-th/9807006}}].


\bibitem{grand}
{\sc O.~Grandjean},
``Non-commutative differential geometry",
Ph.D. thesis, Eidgen\"ossische Technische Hochschule, Z\"urich, 1997.


\bibitem{madore}
{\sc J.~Madore},
{\em An Introduction to Noncommutative Differential Geometry and its Physical
Applications}, London Mathematical Society Lecture Note Series, vol. 257,
Cambridge University Press, 1995.


\bibitem{gratus}
{\sc J.~Gratus},
``Non-commutative differential geometry, and the matrix representations of
generalised algebras",
{\em J.\ Geom.\ Phys.\/} {\bf 25} (1998) 227--244
[\href{http://arxiv.org/abs/q-alg/9703034}{{\tt q-alg/9703034}}].


\bibitem{torel}
{\sc V.C.~de Andrade and J.G.~Pereira},
``Torsion and the electromagnetic field",
{\em Int.\ J.\ Mod.\ Phys.} {\bf D8} (1999) 141--151
[\href{http://arXiv.org/abs/gr-qc/9708051}{{\tt gr-qc/9708051}}].


\bibitem{fradtsey}
{\sc E.~S.~Fradkin and A.~A.~Tseytlin},
``Effective field theory from quantized strings",
{\em Phys.\ Lett.\/} {\bf B158} (1985) 316--322;
``Effective action approach to superstring theory",
{\em Phys.\ Lett.\/} {\bf B160} (1985) 69--76;
``Nonlinear electrodynamics from quantized strings",
{\em Phys.\ Lett.\/} {\bf B163} (1985) 123--130;
``Quantum string theory effective action",
{\em Nucl.\ Phys.\/} {\bf B261} (1985) 1--27.


\bibitem{friedan}
{\sc D.H.~Friedan},
``Nonlinear Models in $\,2+\ep\,$ Dimensions",
Ph.D. thesis, Lawrence Berkeley Laboratory, Berkeley, 1980,
{\em Annals\ Phys.\/} {\bf 163} (1985) 318--419.


\bibitem{princ}
{\sc K.~Gaw\c{e}dzki},
``Lectures on conformal field theory",
1996-97 Quantum Field Theory program at IAS, Princeton,
available at \href{http://www.math.ias.edu/QFT/fall/index.html}{{\tt http://www.math.ias.edu/QFT/fall/index.html}}.


\bibitem{V}
{\sc G.A.~Vilkovisky},
``The unique effective action in quantum field theory",
{\em Nucl.\ Phys.\/} {\bf B234} (1984) 125--137.


\bibitem{BarV}
{\sc A.O.~Barvinsky and G.A.~Vilkovisky},
``The generalized Schwinger--DeWitt technique in gauge theories and quantum
gravity",
{\em Phys.\ Rep.\/}  {\bf 119} (1985) 1--74.


\bibitem{dW}
{\sc B.S.~DeWitt},
``The Effective Action",
in {\em Quantum Field Theory and Quantum Statistics: Quantum Statistics and
Methods of Field Theory}, I.A.~Batalin, C.J.~Isham, and G.A.~Vilkovisky (eds.),
Hilger, 1987.


\bibitem{jpstein}
{\sc J.~Pawe\l czyk and H.~Steinacker}, 
``Matrix description of D-branes on 3-spheres",
{\em JHEP\/} {\bf 0112} (2001) 018 
[\href{http://arXiv.org/abs/hep-th/0107265}{{\tt hep-th/0107265}}];
``A quantum algebraic description of D-branes on group manifolds", 
{\em Nucl.\ Phys.\/} {\bf B638} (2002) 433--458
[\href{http://arXiv.org/abs/hep-th/0203110}{{\tt hep-th/0203110}}].


\bibitem{fusion}
{\sc J.~Pawe\l czyk and R.R. Suszek}, 
``Brane bulk couplings and condensation from REA fusion", 
{\em JHEP\/} {\bf 0604} (2006) 009 
[\href{http://arXiv.org/abs/hep-th/0503240}{{\tt hep-th/0503240}}].

 
\bibitem{ydri}
{\sc P.~Castro-Villarreal, R.~Delgadillo-Blando, and B.~Ydri},
``A gauge invariant UV-IR mixing and the corresponding phase transition for
$\,U(1)\,$ fields on the fuzzy sphere",
{\em Nucl.\ Phys.\/} {\bf B704} (2005) 111--153
[\href{http://arXiv.org/abs/hep-th/0405201}{{\tt hep-th/0405201}}].


\bibitem{scale}
{\sc C.-S.~Chu, J.~Madore, and H.~Steinacker},
``Scaling limits of the fuzzy sphere at one loop",
{\em JHEP\/} {\bf 0108} (2001) 038
[\href{http://arXiv.org/abs/hep-th/0106205}{{\tt hep-th/0106205}}].


\bibitem{kost}
{\sc B.~Kostant},
``A cubic Dirac operator and the emergence of Euler number multiplets of
representations for equal rank subgroups",
{\em Duke\ Math.\ J.\/} {\bf 100} (1999) 447--501.


\bibitem{brink}
{\sc L.~Brink and P.~Ramond},
``Dirac equations, light cone supersymmetry, and superconformal algebras''
[\href{http://arXiv.org/abs/hep-th/9908208}{{\tt hep-th/9908208}}].


\bibitem{CDS}
{\sc A.~Connes, M.R.~Douglas, and A.~Schwarz},
``Noncommutative geometry and Matrix theory: compactification on tori",
{\em JHEP\/} {\bf 9802} (1998) 003
[\href{http://arXiv.org/abs/hep-th/9711162}{{\tt hep-th/9711162}}].


\bibitem{seiwitt-nc}
{\sc N.~Seiberg and E.~Witten},
``String theory and non-commutative geometry",
{\em JHEP\/} {\bf 9909} (1999) 032
[\href{http://arXiv.org/abs/hep-th/9908142}{{\tt hep-th/9908142}}].


\bibitem{sch}
{\sc A.~Schwarz}, 
``Morita equivalence and duality",
{\em Nucl.\ Phys.\/} {\bf B534} (1998) 720--738 
[\href{http://arXiv.org/abs/hep-th/9805034}{{\tt hep-th/9805034}}].


\bibitem{bmz}
{\sc D.~Brace, B.~Morariu, and B.~Zumino}, 
``Dualities of the Matrix Model from T-duality of the type II string", 
{\em Nucl.\ Phys.\/} {\bf B545} (1999) 192--216
[\href{http://arXiv.org/abs/hep-th/9810099}{{\tt hep-th/9810099}}]; 
"T-duality and Ramond--Ramond backgrounds in the Matrix Model", 
{\em Nucl.\ Phys.\/} {\bf B549} (1999) 181--193
[\href{http://arXiv.org/abs/hep-th/9811213}{{\tt hep-th/9811213}}].


\bibitem{ksch}
{\sc A.~Konechny and A.~Schwarz}, 
``BPS states on noncommutative tori and duality",\nl
{\em Nucl.\ Phys.\/} {\bf B550} (1999) 561--584
[\href{http://arXiv.org/abs/hep-th/9811159}{{\tt hep-th/9811159}}].


\bibitem{psch}
{\sc B.~Pioline and A.~Schwarz}, 
"Morita equivalence and T-duality (or $\,B\,$ versus $\,\Theta\,$)", 
{\em JHEP\/} {\bf 9908} (1999) 021 
[\href{http://arXiv.org/abs/hep-th/9908019}{{\tt hep-th/9908019}}].

\end{thebibliography}
\end{document}